\PassOptionsToPackage{pdfpagelabels=false}{hyperref}
\documentclass{aa}
\usepackage[varg]{txfonts}
\usepackage{graphicx}
\usepackage{newtxtext,newtxmath}
\usepackage{xspace}
\usepackage{xcolor}
\usepackage{tabularx}
\usepackage{bm}
\RequirePackage[colorlinks=true,linkcolor=blue,citecolor=blue,urlcolor=blue]{hyperref}
\usepackage[autostyle=true, german=quotes, english=american]{csquotes}
\MakeOuterQuote{"}

\newcommand{\LambdaCDM}{$\Lambda\rm CDM$}
\newcommand{\metalline}[2]{#1\,\textsc{#2}}
\newcommand{\nth}[1]{#1\textsuperscript{th}}

\begin{document} 

\title{The abundance and origin of cool gas in galaxy clusters \\in the TNG-Cluster simulation}
\titlerunning{The origin of cool gas in TNG-Cluster}

\author{Milan Staffehl\inst{1}\thanks{milan.staffehl@stud.uni-heidelberg.de}
      \and Dylan Nelson\inst{1}
      \and Mohammadreza Ayromlou\inst{2,1}
      \and Eric Rohr\inst{3,4}
      \and Annalisa Pillepich\inst{3}
}

\institute{Institut f\"ur theoretische Astrophysik (ITA), University of Heidelberg, Albert-Ueberle-Straße 2, D-69120 Heidelberg \label{1}
\and Argelander Institute f\"ur Astronomie, Auf dem H\"ugel 71, D-53121 Bonn, Germany \label{2}
\and Max-Planck-Institut f{\"u}r Astronomie, K{\"o}nigstuhl 17, D-69117 Heidelberg, Germany \label{3}
\and Universit\"{a}t Heidelberg, Exzellenz-Cluster STRUCTURES, Philosophenweg 12, 69120 Heidelberg, Germany \label{4}
}

\date{}

\abstract{In addition to the hot intracluster medium, massive galaxy clusters host complex, multi-phase gaseous halos. In this work, we quantify the abundance, spatial distribution, and origin of the cool ($T\leq10^{4.5}\,\rm{K}$) gas within clusters. To do so, we combine the TNG-Cluster and TNG300 cosmological magnetohydrodynamical simulations, yielding a sample of 632 simulated galaxy clusters at $z=0$ with masses $M_{200c} \sim 10^{14-15.4}\,\rm{M_\odot}$. We find that cool gas is present in every cluster at $z=0$, although it constitutes only a small fraction of the total gas mass within twice the virial radius, ranging from $\sim 10^{-4}$ to a few per cent. The majority of cool gas resides in the cluster outskirts in infalling satellites and other halos. More rarely, cool gas can also be present in the central regions of clusters. More massive halos contain larger amounts (but not fractions) of cool gas ($\sim 10^{10-12}\,\rm{M_\odot}$), and we identify correlations between cluster cool gas fraction and several global halo and galaxy properties at fixed halo mass. Using Monte-Carlo Lagrangian tracer particles, we then track the origin of cool gas in present-day clusters. We find that the primary source is recent accretion at $z \lesssim 0.1$, predominantly in the form of pre-cooled gas carried by infalling satellite galaxies and other halos. However, in-situ cooling of the hot intracluster medium gas accreted at earlier epochs also contributes, especially in present-day cool-core clusters.}


\maketitle


\section{Introduction}
\label{sec:introduction}

In the standard model of structure formation within $\Lambda$ Cold Dark Matter (\LambdaCDM), halos form from the gravitational collapse of dark matter and subsequent hierarchical mergers. Halos then accrete gas that cools to ultimately form stars and galaxies \citep{ReesOstriker1977, Silk1977, WhiteRees1978}. The continual mergers of galaxies leads to increasingly massive systems, and galaxy clusters are the end result of this process of hierarchical structure formation. 

While galaxy clusters have an abundance of galaxies, the majority of the baryonic mass of a galaxy cluster is not locked in stars, but remains, and is observed, in a reservoir of gas: the intracluster medium (ICM). This ICM is typically at the virial temperature of the cluster, reaching $10^{7-8} \,\rm K$ and above \citep{Voit2005ReviewOfClusters}. According to models, clusters continually accrete more gas from the surrounding intergalactic medium (IGM) through the filaments of the cosmic web \citep{Zinger2016Accretion}. 

Due to its extreme temperatures, the ICM is visible in the X-ray regime chiefly via thermal bremsstrahlung emission \citep{Sarazin1986XRay, JonesForman1999XRayEinstein}, enabling an observational probe of cluster hot gas. Interestingly, however, observations at longer wavelengths reveal that the ICM also contains cold \citep[$T < 10^4\, \rm K$; e.g.][]{Salome2006,Russell2016,McNamara2014,Omoruyi2024}, cool \citep[$T \sim 10^4\, \rm K$; e.g.][and below]{Olivares2019CoolFilaments,Olivares2025HalphaFilaments} and warm-hot \citep[$T \sim 10^{5-6}\, \rm K$, e.g. ][]{Werner2008,Burchett2018WarmHot} gas phases.

Observationally, the cool phase of the ICM has been seen both in emission and absorption. Emission-based observations have mostly targeted individual systems, revealing for example the 90-kpc extended H-alpha filaments in NGC 1275 at the center of Perseus \citep{Fabian2003PerseusFilaments, Rhea2025} as well as blue star-forming gas clumps and filaments in Virgo \citep{Dey2024BlueClumps}. At higher-redshifts $z\gtrsim2$, Ly$\alpha$ nebulae and even radio observations of CO have revealed multiphase gas in protoclusters \citep{Steidel2000,Matsuda2012,Chen2024}.

Absorption with background quasar spectra allows the characterization of cool gas across many systems. A prominent tracer is \metalline{Mg}{ii}, which suggests significant amounts of cool gas in galaxy clusters \citep{Fresco2024ColdGasMgII}. While the detection rate of \metalline{Mg}{ii} absorbers is higher within the cluster than outside it ($3-5$ virial radii), there are fewer absorbers per galaxy in clusters than in the field \citep{Lee2021MgIIAbsorbers}. Nonetheless, \metalline{Mg}{ii} absorbers are found in both the cluster interior as well as its outskirts: \cite{Mishra2022Discovery} find strong \metalline{Mg}{ii} (and marginal \metalline{Fe}{ii}) absorption in the outskirts of $z \approx 0.5$ clusters, which persists even when excluding quasar-cluster pairs probing the circumgalactic medium (CGM) of satellite galaxies. This indicates the presence of metal-rich, cool gas in the ICM in cluster outskirts. However, contributions of the interstellar medium (ISM) or CGM of low-mass galaxies to the absorption cannot be ruled out. Similarly, \cite{Anand2022Cool} observe significant covering fractions of \metalline{Mg}{ii} absorbers within $R_{500}$ of clusters. Since no correlation between \metalline{Mg}{ii} strength and star formation rate of the nearest galaxy is observed, they conclude that the cool gas traced by the absorbers is dominated by cool gas clouds in the ICM rather than the ISM of the galaxies.

Another prominent tracer for cool gas is atomic hydrogen (\metalline{H}{i}). Cluster galaxies exhibit reduced covering fractions of \metalline{H}{i} absorption compared to field galaxies, with the ISM/CGM of these galaxies being increasingly \metalline{H}{i} depleted with increasing environmental density towards the cluster center \citep{Burchett2018WarmHot}. Studies of individual nearby galaxy clusters such as the Coma and Fornax cluster using \metalline{H}{i} detection have similarly found evidence of cool gas in these clusters, which is prominently stripped from infalling galaxies \citep{Molnar2022Westerbork}. Dwarf galaxies in the Fornax cluster show signs of \metalline{H}{i} removal due to ram-pressure stripping \citep{GunnGott1972RPS}, while the \metalline{H}{i} detected dwarfs also avoid the densest region of the cluster, suggesting that \metalline{H}{i} gas is stripped from infalling dwarfs in the cluster interior \citep{Kleiner2023MeerKAT}.

From a theoretical perspective, cool gas may well be transported into a cluster via infalling galaxies that become cluster members. These galaxies undergo ram-pressure stripping \citep{Zinger2016RPS, Ayromlou2019New,Yun2019Jellyfish}, a process that is expected to occur not only within cluster boundaries but also extend out to several megaparsecs \citep{Bahe2013RPS,Wetzel2014Galaxy,Ayromlou2021Galaxy, Zinger2024Jellyfish}. Clear evidence of ongoing gas loss is provided by the morphology of jellyfish galaxies, which have been observed across wavelengths \citep[e.g.][]{Ebeling2014Jellyfish} and which exhibit visible tails of stripped gas \citep{Poggianti2017Jellyfish}. The infall of galaxies into clusters can also compress the gas in galaxy disks via ram-pressure \citep{Lee2017RamPressure}, facilitating further cooling \citep{Nelson2020Resolving} and star formation \citep{Leee2020RPSDualEffects, Goeller2023Jellyfish}. Additionally, tidal interactions and collisions \citep{SpitzerBaade1951Collisions, Moore1996Harassment, Bekki1998TidalStripping} have been both theorized and observed to impact cluster satellite galaxies. Collectively, these processes deprive cluster galaxies of their cool gas, chiefly determining their evolution, but also depositing such cool gas into the ICM and potentially contributing to its multi-phase nature \citep{Rohr2023Jellyfish,Rohr2024HotSatelliteCGM}.

In addition to satellites and gas accretion from the IGM, energetic outflows from the central supermassive black hole (SMBH) in clusters may also contribute. AGN feedback is observed in a diversity of galaxy clusters \citep[e.g.][]{McNamara2007AGNFeedback, Fabian2012AGNFeedback}. Simulations suggest that AGN-driven outflows can redistribute gas from small to large scales \citep{Nelson2019TNG50}, and even beyond the halo entirely \citep{Ayromlou2023Feedback,Sorini2022}. These outflows may also transfer cold gas from the central galaxy to the ICM, one case where magnetic fields and thermal conduction may impact the survival of such cool gas \citep{Fournier2024,Hong2024ColdStreams,Talbot2024}. In addition, the central galaxies of clusters can potentially support high star formation rates, if cool gas is sufficiently abundant in their centers \citep{Bassini2020,Nelson2024TNGCluster}, i.e. if the cluster is in a cool-core state \citep{Lehle2024CoolCores}.

Finally, the hot ICM can also cool in-situ. Given a short enough cooling time $t_{\rm cool}$, density perturbations in the ICM may lead to the cooling of hot gas \citep{Sharma2012Multiphase}, which can then accrete onto the central galaxy of the cluster, a process referred to as condensation or precipitation \citep{Voit2015Precipitation, Voit2017Precipitation}. Theoretically, the broad question of whether such cool gas can survive within a hot background remains unclear \citep[e.g.,][]{Sparre2020ColdCloud,Fielding2022MultiphaseWinds}. Furthermore, the details of how and if cluster gas can rapidly cool may be intricately linked with AGN feedback and thus activity in the central galaxy \citep{Li2015,Beckmann2019}.

In order to disentangle the many possible origins and contributors to cool gas in clusters -- satellites, accretion, outflows, and in-situ cooling -- we turn to numerical simulations. Modern cosmological simulations of galaxies produce multiphase gaseous halos \citep{Emerick2015Warm,Nelson2020Resolving,Heintz2024NeutralGasProtoCluster}. These models enable us to assess the abundance and origin of cool gas in galaxy clusters by accounting for both their diffuse ICM as well as the gas in and around cluster galaxy members.

In this work we investigate the abundance, spatial distribution, and physical origin of cool gas in $z=0$ galaxy clusters. We quantify the demographics of the cool cluster gas today and track its origin over the past $\approx13$~billion~years. To do so, we combine the new TNG-Cluster cosmological magnetohydrodynamical simulation \citep{Nelson2024TNGCluster} with the existing cluster sample from TNG300\footnote{\url{www.tng-project.org}}. This allows us to study a large sample of $\sim 630$ clusters across the full mass range at $z=0$, $10^{14}\leq M_{200c}/M_{\odot}\lesssim 10^{15.4}$.

Both simulations use the fiducial TNG galaxy formation model, and this has been validated in particular for cluster galaxy populations, showing that they are in reasonable agreement in terms of star-formation activity, neutral hydrogen content, quenched fractions, and overall radial distribution \citep[e.g.,][]{Nelson2018TNGPublicationI,Stevens2018NeutralHydrogen, Donnari2021QuenchedFractionsComparisons,Stevens2021MolecularHydrogen,Riggs2022}. Of particular note, \citet{Rohr2024CoolerPastClusters} recently showed that TNG-Cluster halos at $z=0$ contain reservoirs of cool gas in their diffuse ICM, i.e. specifically excising satellites and hence their cool gas contribution. Further, they showed that cluster progenitors at $z\approx2$ had more cool gas in their ICM than their descendants today, and the onset of kinetic SMBH feedback drives this decrease in cool ICM mass. Moreover, at a fixed halo mass, the $z=0$ cool ICM mass in TNG-Cluster correlates with the number of gaseous satellites, suggesting that the cool ICM owes its existence at least in part to the passage of satellites \citep{Rohr2024CoolerPastClusters}.

This paper is structured as follows: Section~\ref{sec:methods} introduces the simulations and our methodology. Section~\ref{sec:results} presents our main results. We first consider cool gas across halos of all masses (Section~\ref{ssc:results:cool_gas_in_massive_halos}), and then examine relationships between cluster cool gas and galaxy and halo properties (Section~\ref{ssc:results:cool_gas_in_galaxy_clusters}), and quantify the spatial and kinematical properties of cool gas (Sections \ref{ssc:results:radial_distribution} and \ref{ssc:results:velocity_distribution}). Moving to origin, we study the assembly history (Section~\ref{ssc:results:the_origin_of_cool_cluster_gas}), temperature evolution and time of cooling (Section~\ref{ssc:results:temperature_development}), and key accretion channels (Section~\ref{ssc:results:structural_origin_of_cool_gas}) of cool gas. We summarize our findings in Section~\ref{sec:conclusions}.


\section{Methods}
\label{sec:methods}

\subsection{The IllustrisTNG and TNG-Cluster Simulations}
\label{ssc:methods:the_tng_simulations}

The original IllustrisTNG project \citep[TNG hereafter; see][]{Nelson2019TNGRelease} consists of three suites of hydrodynamical simulations over three different volumes: TNG50 \citep{Pillepich2019TNG50,Nelson2019TNG50}, TNG100, and TNG300 \citep{Nelson2018TNGPublicationI, Springel2018TNGPublicationII, Naiman2018TNGPublicationIII, Pillepich2018TNGPublicationIV, Marinacci2018TNGPublicationV}. The recent TNG-Cluster simulation \citep{Nelson2024TNGCluster} extends the sample to massive galaxy clusters.

For this work we combine the TNG300 and TNG-Cluster simulations. TNG300 simulates a cosmological box of edge length $302.6 \,\rm Mpc$ (comoving) with $2 \times 2500^3$ baryonic fluid cells and DM particles each, reaching a baryon mass resolution of $\sim 1.1 \times 10^7 \,\rm M_\odot$. The TNG-Cluster simulation consists of 352 zoom-in simulations of massive galaxy clusters selected from a $\sim 1$\.comoving Gpc box. These clusters were chosen based solely on their $z=0$ halo mass, making the selection unbiased with respect to all other properties. The resolution of TNG-Cluster is the same as TNG300, making it easy to combine galaxy and halo samples across these two simulations.

The TNG and TNG-Cluster simulations employ a well-tested model for galaxy formation \citep{Weinberger2017TNG,Pillepich2018TNG} based on the \textsc{arepo} code \citep{Springel2010AREPO,Weinberger2020AREPORelease}. It follows the coupled evolution of stars, supermassive black holes (SMBHs), gas, cold dark matter, and magnetic fields by solving the equations of ideal, continuum magnetohydrodynamics (MHD) on an unstructured moving Voronoi-tessellated grid, using a Godunov finite volume scheme, as well as self-gravity using a TreePM approach.

The following physical processes are modeled:
(i)~primordial and metal-line gas cooling, plus heating by a redshift-dependent UV background radiation field, including self-shielding in the dense ISM;
(ii)~stochastic star formation in the dense ($n \gtrsim 0.1 \,\rm cm^{-3} $) ISM, adopting a Chabrier initial mass function;
(iii)~chemical enrichment of the ISM via supernovae of type Ia and II, and winds from AGB stars, including the abundance of nine individual elements (H, He, C, N, O, Ne, Mg, Si and Fe) as well as overall metallicity \citep[see][]{Pillepich2018TNG};
(iv)~galactic-scale stellar winds, driven by supernovae type\,II;
(v)~formation, merging, and growth of SMBHs \citep[see][]{Weinberger2017TNG}; 
(vi)~SMBH feedback, injecting energy in the form of discrete, directed kinetic energy injections in random directions at low-accretion rates (kinetic mode), and continuous isotropic thermal energy injection at high-accretion rates (quasar mode). Radiation from AGN also modulates the cooling of nearby gas. 

All TNG simulations, including TNG-Cluster, save 100 simulation snapshots from $z \approx 20$ to $z = 0$, and adopt a cosmology consistent with the results from the \cite{Planck2016Cosmology}, that is: $\Omega_{\Lambda,0} = 0.6911$, $\Omega_{\rm m, 0} = 0.3089$, $\Omega_{\rm b, 0} = 0.0486$, $\sigma_8 = 0.8159$, $n_{\rm s} = 0.9667$, and $h = 0.6774$.

To identify dark matter halos, the friends-of-friends algorithm \citep{Davis1985FoF} identifies clustered dark matter (DM) particles. Each baryonic particle is linked to its nearest DM particle. The \textsc{subfind} algorithm \citep{Springel2001Subfind,Dolag2009Subfind} then identifies gravitationally-bound substructures of these FoF groups, denoted `subhalos'. Subhalos containing stars and/or gas particles are referred to as galaxies. We use the terms `FoF group' and `halo' interchangeably. The position of a halo is the location of the particle/cell with the minimum gravitational potential energy, within its central i.e. most massive galaxy. All other subhalos are satellites. The \textsc{SubLink} algorithm \citep{RodriguezGomez2015Sublink} records the merger history of subhalos identified by \textsc{subfind} by constructing merger trees. The first (or main) progenitor is the progenitor with the \enquote{most massive history} (see \citealp{DeLuciaBlaizot2007MergerTree} for a detailed definition), and this series of subhalos through time is the main progenitor branch.

In order to trace the origins of gas in the simulation volume, we use Monte-Carlo Lagrangian tracer particles \citep{Genel2013Tracers,Nelson2013}. Each tracer represents a constant mass, equal to the initial mass of a gas cell. The tracers follow the mass flow of the fluid probabilistically, such that they statistically reproduce its mass distribution. This lets us trace the Lagrangian evolution and past evolution of fluid elements. 

\subsection{Cluster sample and definitions}
\label{ssc:methods:analysis_choices}

We consider all halos with masses $M_{200c} \geq 10^{14.0}\,\rm M_\odot$ at redshift $z = 0$ as clusters. Across the TNG300 and TNG-Cluster simulations, there are a total of 632 clusters — 280 in TNG300 and 352 in TNG-Cluster. Throughout this work, we refer to the halo mass as $M_{200c}$, the mass enclosed by the radius $R_{200c}$ such that the total average enclose density is equal to 200 times the critical density of the universe at that time. We also use the terms virial mass and virial radius to refer to $M_{200c}$ and $R_{200c}$, respectively.

The virial temperature of a halo is calculated following \cite{BarkanaLoeb2001VirialTemperature} as $T_{\rm 200c} = \mu m_p V_{\rm 200c}^2 / (2 k_{\rm B})$ where $V_{\rm 200c} = (GM_{\rm 200c} / R_{\rm 200c})^{1/2}$. Here $\mu = 0.6$ is the mean molecular weight of the gas, $m_p$ the proton mass, $k_{\rm B}$ the Boltzmann constant, $G$ the gravitational constant, and $M_{200c}$ and $R_{200c}$ the virial mass and virial radius respectively. Since gas cannot cool below the effective temperature floor of $\sim 10^4 \,\rm K$ in the TNG model, we set star-forming gas to a temperature of $10^3 \,\rm K$, consistent with its cold and mass-dominant phase \citep{Springel2003}. We then adopt a temperature division to separate gas into three phases: cool gas with $T \leq 10^{4.5} \,\rm K$ \citep[as in][]{Rohr2024CoolerPastClusters}, warm gas with $10^{4.5} < T \leq 10^{5.5} \,\rm K$, and hot gas with $T > 10^{5.5} \,\rm K$. This fixed definition of cool gas is applied throughout this entire analysis, and at all redshifts. The $z=0$ clusters with total mass $\sim10^{14-15.4}\,\rm{M_\odot}$ have virial temperatures $\sim10^{7-8}$~K, in the host phase regime (see below for more details).

We associate all gas within $2R_{200c}$ with each cluster, irrespective of halo membership. The sole exception to this is Section~\ref{ssc:results:cool_gas_in_massive_halos} (and Figures~\ref{fig:results:mass_trends_tng300} and \ref{fig:results:temperature_distribution}), where instead we consider only gas in each FoF halo, and a list of cluster properties listed in Table \ref{tb:results:cluster_properties} which is similarly limited to FoF gas. In both cases, and differently than in \citet{Rohr2024CoolerPastClusters}, we consider throughout this paper all cluster gas, i.e. including, and without distinguishing between, 
the smooth ICM and gas that belongs to the central and satellite galaxies as well as to other halos.

\subsection{Tracing cool gas and origin categories}
\label{ssc:methods:analysis_tracing}

To follow the origin of cool gas, we focus on TNG-Cluster only, where all halos have identified merger trees for redshifts $0 \leq z \leq 8$, to which we limit our analysis. 

At redshift zero, for each selected cluster we identify all cool gas cells ($T < 10^{4.5}\,\rm K$) within $2R_{200c}$ from the cluster center. For each of these cells, all associated tracer particles are identified, and traced back to redshift $z = 8$. These selected tracers have parent gas cells at each snapshot that provide the predecessor gas properties of the $z=0$ cool gas. As each tracer particles represents a mass element of the underlying fluid, we weight all properties of gas cells by the total mass of the selected tracers they contain. 

Although all our tracers have gas cell parents at $z=0$, they can also reside in stellar particles at earlier times. Stellar mass loss processes transfer tracers from stars to gas. At all redshifts, the fraction of our selected tracers in stars is low—typically below 2\% and never above 6\%. Since stars do not have defined temperature or density, we exclude tracers in stellar particle parents from any analysis involving these properties. However, because position and velocity are available for all particle types, we include tracers in stars for those analyses. SMBH particles are not included in our analysis, as they do not return mass to the environment and therefore cannot contribute to cool gas at $z=0$.

Every particle or cell in the simulation has at most one uniquely identified parent halo and parent subhalo, and the host halos and subhalos of tracers are defined similarly. Particles and tracers may belong to a FoF group while not belonging to any subhalo, but never the other way around. At early times, gas and tracers often reside outside any subhalo or halo. Therefore, for the cool gas that we identify at $z=0$ in each studied cluster, we define the following four categories for analyzing its origin via the tracers:

\begin{enumerate}[(i)] 
    \item \textbf{Unbound:} Tracers that have neither a parent halo nor a parent subhalo; in other words, they reside in the intergalactic medium (IGM). 
    \item \textbf{Other halo:} Tracers that have a parent halo that is not the main progenitor of the $z=0$ cluster under scrutiny — for example, those residing in a different halo at $z>0$ that will merge with the protocluster in the future. 
    \item \textbf{Primary halo:} Tracers whose parent halo is the primary cluster halo under scrutiny (or its main progenitor for $z>0$) and that either belong to the primary subhalo of the cluster or have no parent subhalo (often referred to as "inner fuzz"). This category includes both the intracluster medium (ICM) and the central galaxy. 
    \item \textbf{Satellite:} Tracers within the primary cluster halo that reside in a satellite parent subhalo (a satellite galaxy). 
\end{enumerate}

Tracers typically belong to different categories at different times, as their parent halo and subhalo change through cosmological evolution.


\section{Results}
\label{sec:results}

\subsection{Cool gas in halos across the mass range}
\label{ssc:results:cool_gas_in_massive_halos}

\begin{figure}
  \centering
  \includegraphics[width=\columnwidth]{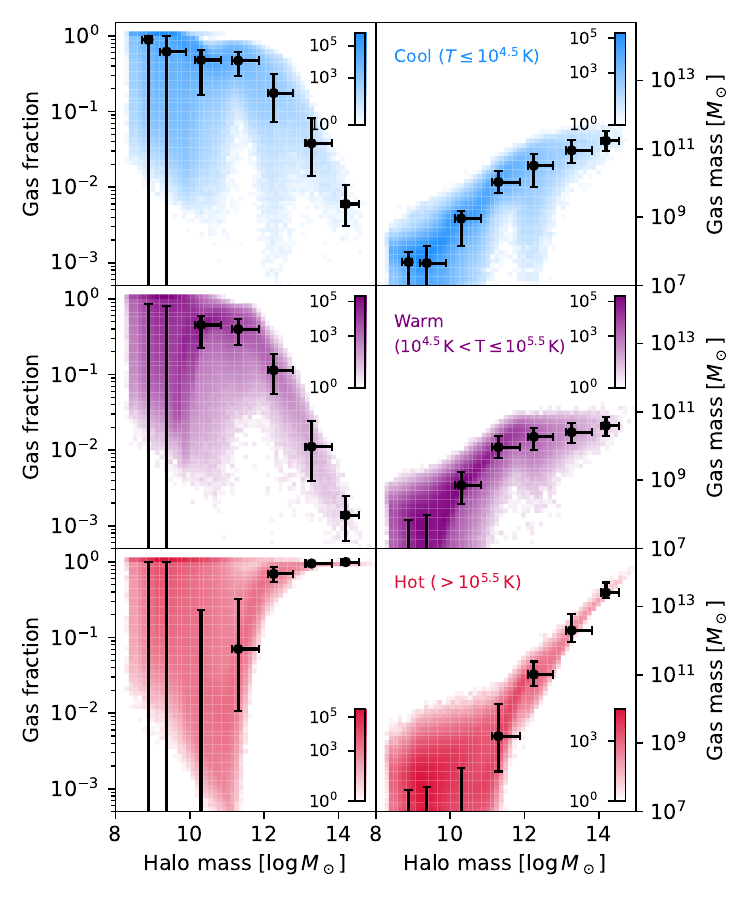}
  \caption{The fraction of each gas phase relative to the total gas mass (left) and the absolute gas mass for each phase (right) as functions of halo mass (x-axis) across all halos and clusters of TNG300 at $z=0$ . We divide gas into cool (first row, blue), warm (second row, purple), and hot (third row, red) components. The color indicates the number of halos per pixel. The black markers are the median in both gas fraction/gas mass and halo mass for mass bins of width 1 dex from $10^8$ to $10^{15} \,\rm M_\odot$. The error bars denote the \nth{16} and \nth{84} percentile along both axes. More massive clusters host larger amounts of cool gas within their FoF halos, even though fractionally the contribution of cool and warm gas to clusters decreases with total mass.}
  \label{fig:results:mass_trends_tng300}
\end{figure}

To contextualize the following findings for galaxy clusters, we begin with a global census of gas in halos as a function of phase.

Figure~\ref{fig:results:mass_trends_tng300} shows the current ($z = 0$) gas content of halos as a function of halo mass. We split gas into three temperature regimes: cool gas ($T \leq 10^{4.5} \,\rm K$, first row) in blue, warm gas ($10^{4.5}\,\rm K < T \leq 10^{5.5}\,\rm K$, second row) in purple, and hot gas ($T > 10^{5.5} \,\rm K$, third row) in red. Here, we consider all FoF gas to comprise the halo gas, regardless of cluster-centric distance. The left column shows the gas mass fraction, normalizing by the total gas mass of the FoF halo. The right column instead shows the total mass of FoF gas in the corresponding temperature range. The black markers give median and 16-84th percentile gas fractions for 1 dex bins in halo mass.\footnote{The significant scatter at low mass is artificially boosted due to the mass resolution of TNG300, i.e. halos in the lowest mass bins do not contain many gas cells.}

Cool gas fraction steadily decreases with halo mass, and drops sharply beyond the group mass scale ($M_{200c} > 10^{13} \,\rm M_\odot$), to only a few per-cent or less in the cluster regime ($M_{200c} > 10^{14} \,\rm M_\odot$). The cool gas mass on the other hand shows a steady increase with increasing halo mass, reaching cool gas masses in the order of $\sim 10^{11} \,\rm M_\odot$ in the cluster mass range.

This value is significantly larger than that measured by \citet[][cool cluster gas masses of $\sim10^{9.5}\,\rm{M_\odot}$]{Rohr2024CoolerPastClusters}, because they consider only FoF gas within the aperture $[0.15,\ 1.0]R_{200c}$ and exclude satellite gas from their analysis, suggesting that either a majority of cool cluster gas today is found at large distances $>R_{200c}$ and/or within satellite galaxies (see below for more details). A noticeable feature in both the cool gas fraction and cool gas mass is the scatter towards lower gas fractions/gas masses around $\sim 10^{12} \,\rm M_\odot$, a scale at which massive galaxies can produce significant AGN feedback, depleting some of the halos of their cool gas by ejection or heating \citep{Zinger2020SMBHFeedback}.

The warm gas behaves similarly to the cool gas, with a steadily decreasing fraction towards higher halo masses, reaching even lower fractions ($\sim 0.1\,\%$) in the cluster regime, and with an increasing total warm gas mass up to around $\sim 10^{11} \,\rm M_\odot$, after which the distribution flattens notably, indicating that in the most massive halos the total warm gas mass does not depend as strongly on halo mass as for the cool gas. Lastly, the hot gas fraction increases with halo mass, as expected given our constant temperature thresholds applied to the evolving virial temperature of dark matter halos. Hot gas fractions reach unity for clusters, that are typically dominated by virialized, X-ray emitting gas.

\begin{figure}
    \centering
    \includegraphics[width=0.47\textwidth]{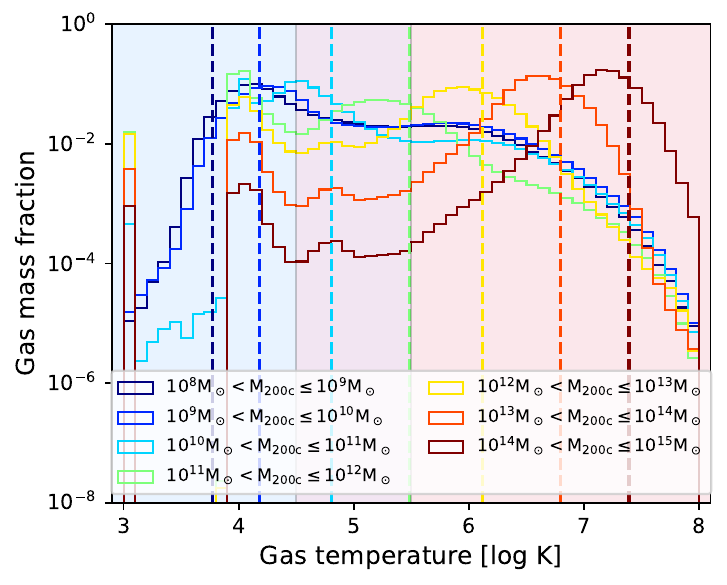}
    \caption{Temperature distributions of gas in halos of TNG300, from $10^8$ to $10^{15} \,\rm M_\odot$ in mass bins of width 1\,dex. The dotted lines show the average virial temperature in the corresponding mass bin. The shaded regions show the temperature regimes in the same colors as Figure~\ref{fig:results:mass_trends_tng300}. Even for the very massive systems that are the focus of this paper (clusters with $> 10^{14} \,\rm M_\odot$, brown curve), a non negligible amount of cool gas resides within the ICM and cluster galaxies.}
    \label{fig:results:temperature_distribution}
\end{figure}

Figure~\ref{fig:results:temperature_distribution} shows the distribution of temperatures of all gas in all halos of TNG300 at $z=0$, in seven mass bins of 1\,dex. For halos of mass $M_{200c} \sim 10^{12-15}\,\rm{M_\odot}$, three peaks can be identified in the distribution, corresponding to the minima of the cooling function for hydrogen and helium (at $\sim 10^4 \,\rm K$ and $\sim 10^{4.8} \,\rm K$ respectively) and the halo virial temperatures (at $\sim 10^{6-7.5} \,\rm K$), that are marked by a colored dashed line. The shaded regions show the temperature regimes in the same colors as Figure~\ref{fig:results:mass_trends_tng300}. All halos have gas in all three phases, where lower mass halos have virial temperatures that typically fall into the warm and cool temperature regimes. As some gas exists orders of magnitude in excess of the corresponding virial temperature, this requires some other non-gravitational source of heating, such as feedback-driven heating. 

For this work, the final mass bin, which includes halos of cluster masses ($M_{200c} > 10^{14} \,\rm M_\odot$), is of primary interest. To enhance the statistical sample from TNG300, we incorporate halos from the TNG-Cluster simulation, allowing for a more robust investigation of the abundance and distribution of cool gas in this mass range.

\begin{figure*}
    \centering
    \includegraphics[width=1.0\textwidth]{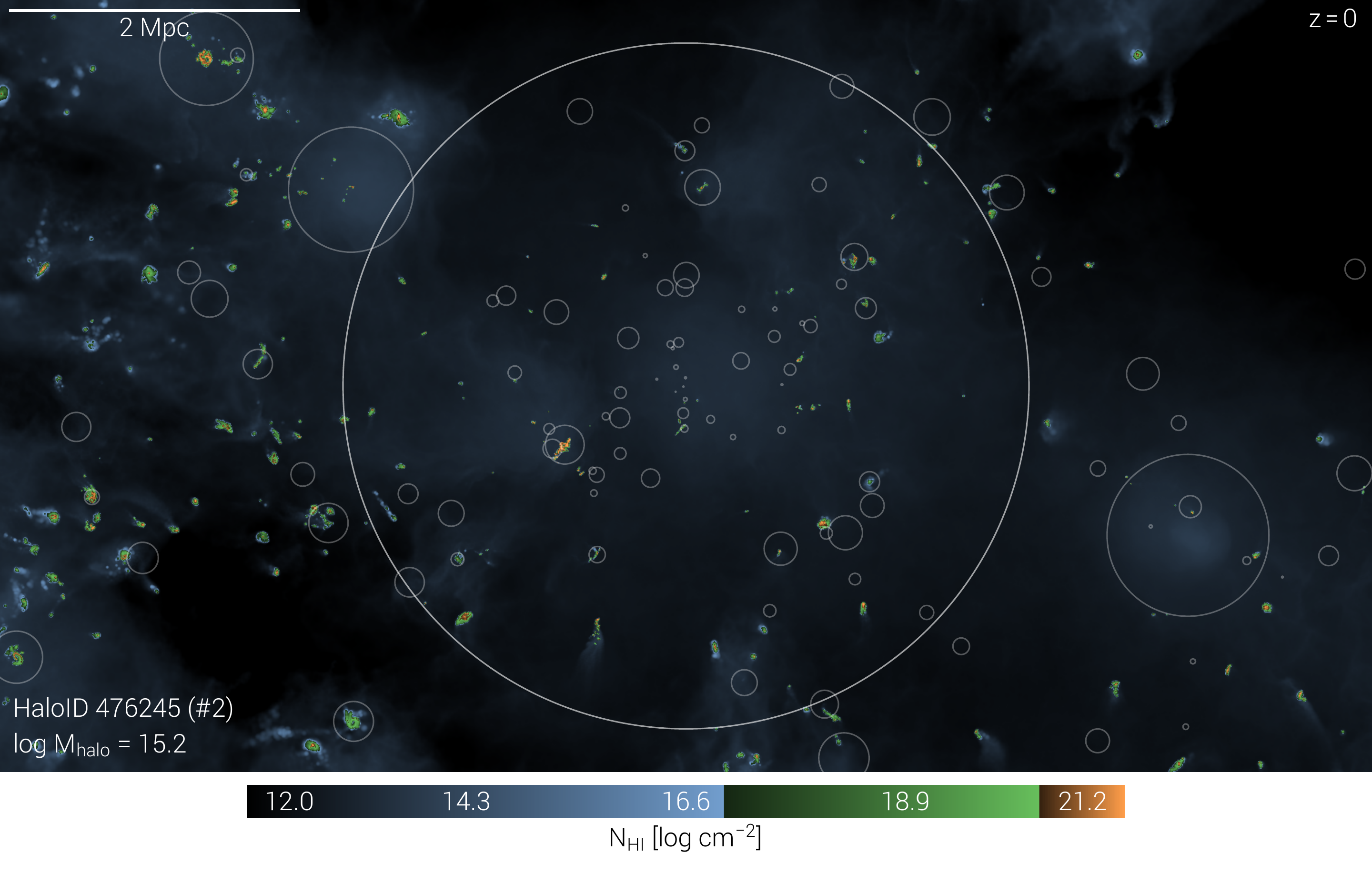}
    \caption{Visualization of the cool gas distribution around the third most massive halo in TNG-Cluster at $z=0$. The field-of-view is $3 R_{\rm 200c}$ from side to side, as well as in the projection direction, which is random with respect to the cluster itself. The large white circle shows the $\sim 2$\,Mpc virial radius of the cluster itself. Color indicates neutral hydrogen column density, a tracer of cool $\sim 10^4$\,K gas. While tracer amounts of cool gas pervade the cluster environment (blue), the vast majority of cool gas mass is at higher column densities (green and orange). This material is predominantly localized in and around satellite subhalos, where the most massive 100 such subhalos are enclosed with white circles, with radii equal to their half mass radii.}
    \label{fig:results:single_halo_vis}
\end{figure*}

\subsection{Cool gas in galaxy clusters}
\label{ssc:results:cool_gas_in_galaxy_clusters}

Figure~\ref{fig:results:single_halo_vis} begins with a visual impression of the distribution of cool gas in and around a massive galaxy cluster. We show the projected neutral HI column density, directly tracing cool $\sim 10^4$\,K gas, on the scale of the virial radius (white circle) of the third most massive TNG-Cluster halo at $z=0$. The color map diverges at $N_{\rm HI} = 10^{17.0}$\,cm$^{-2}$ and $10^{20.3}$\,cm$^{-2}$, reflecting the common definitions for Lyman-limit systems and damped Lyman-alpha absorbers \citep[LLSs and DLAs, respectively;][]{Peroux2020}.

A strong association between cool gas and satellites is apparent. We emphasize this by marking the 100 most massive satellite subhalos of this cluster halo with white circles, each extending to the respective subhalo half mass radii. Indeed, most of the cool gas mass exists at higher column densities (green and orange) that are preferentially found within and near satellites. It is clear, however, that many of the most massive satellites, indicated by the size of the enclosing circle, do not contain large HI reservoirs, due to environmental and/or internal processes. That is, at a fixed a satellite (and host) mass, more recent infallers are more likely to retain more cool gas than other satellites that have been experiencing a larger cumulative ram pressure \citep{Rohr2023Jellyfish,Rohr2024HotSatelliteCGM}. Further, within the TNG model, massive cluster satellites (stellar mass $\gtrsim10^{10.5}\,\rm{M_\odot}$) tend to lose their cool gas (and quench their star-formation) due to internal SMBH feedback \citep{Donnari2021PreProcessing,Rohr2024HotSatelliteCGM}. Thus, it is likely that many of the satellites contributing to the cool cluster gas are recent infallers that have not experienced significant amounts of kinetic SMBH feedback. 

In contrast to the localized high density HI, the diffuse low density material (blue) contains little mass. It also traces the large-scale distribution of satellite galaxies, as well as the preferred directions along which they are accreting from the local large-scale structure, i.e. when outside of the cluster virial radius, preferentially towards the upper left and lower left.

\begin{figure*}
    \centering
    \includegraphics[width=0.48\textwidth]{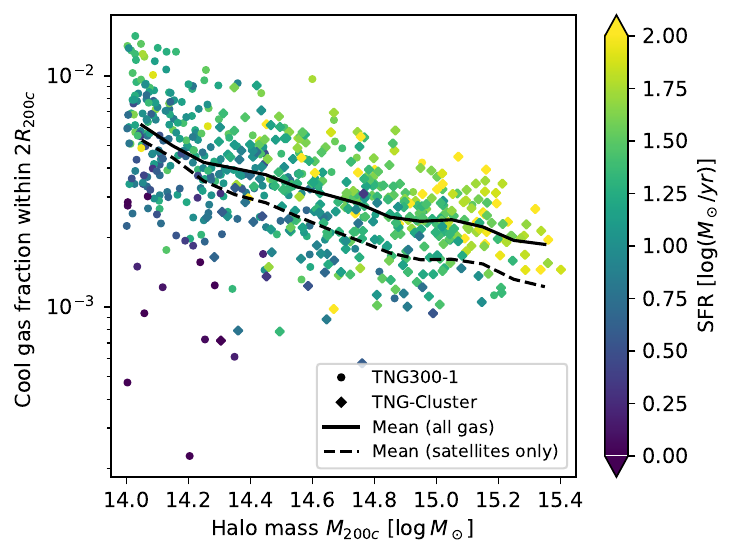}
    \includegraphics[width=0.48\textwidth]{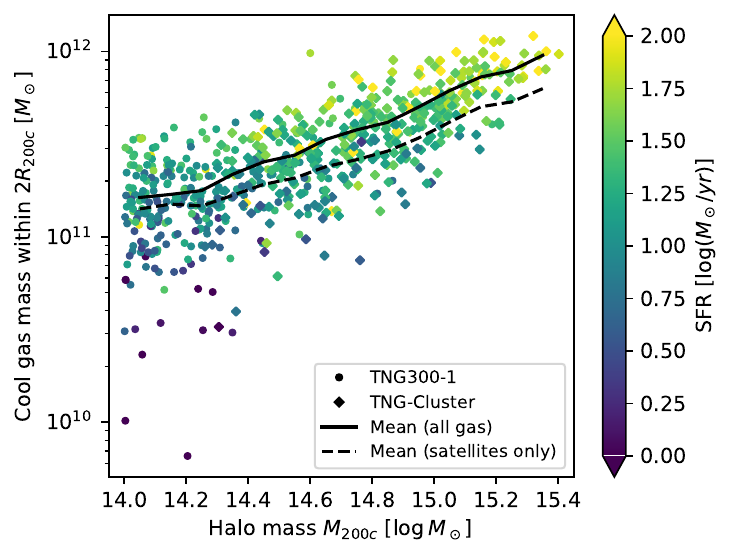}
    \includegraphics[width=\textwidth]{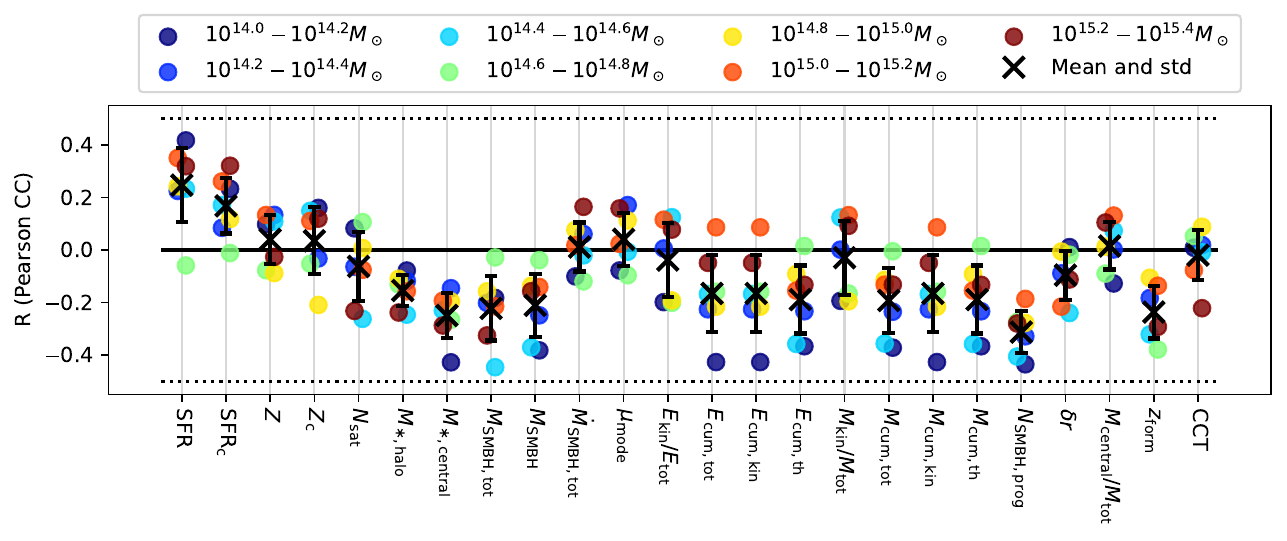}
    \caption{The abundance of cool gas in galaxy clusters at $z=0$ according to the TNG300 and TNG-Cluster simulations. Top: Cool ($T < 10^{4.5} K$) to total gas mass fraction (left) and cool gas mass (right) as a function of cluster mass $M_{200c}$. All gas within $2R_{200c}$ from the cluster center is considered, hence including both ICM as well as satellite gas. The circles (diamonds) show TNG300 (TNG-Cluster) halos at $z=0$, colored by the total star formation rate of the halo, a measure of the ongoing SF activity in both the central and satellite galaxies. The solid black lines show the running average of cool gas mass fraction and cool gas mass respectively, while the dashed black lines show the running average only for gas in satellites within $2R_{200c}$, regardless of host halo. Bottom: Pearson correlation coefficient between cool gas fraction within $2R_{200c}$ and a number of selected cluster properties (listed in detail in Table~\ref{tb:results:cluster_properties}), in seven mass bins of 0.2 dex depicted by colored dots. The points have been given small horizontal shifts for visual clarity. The black crosses mark the mean correlation coefficient over all mass bins, including error bars for standard deviation. A large fraction of cool gas within $2R_{200c}$ is gravitationally bound to satellite galaxies, and its mass fraction weakly correlates with multiple  properties of the cluster.}
    \label{fig:results:cool_gas_vs_cluster_mass}
\end{figure*}

To quantify the population-level trends, Figure~\ref{fig:results:cool_gas_vs_cluster_mass} presents the cool gas fraction and cool gas mass as functions of halo mass for all 632 clusters in the combined TNG300 and TNG-Cluster sample at $z=0$. In contrast to Section~\ref{ssc:results:cool_gas_in_massive_halos}, we now associate all gas within twice the virial radius, $2R_{200c}$, with each cluster (regardless of halo membership). Our results show that the cool gas fraction decreases with increasing cluster mass, ranging from $\sim10^{-2}$ at a mass of $10^{14}\,\rm{M_\odot}$ down to $\approx2 \times 10^{-4}$ for the most massive clusters. Meanwhile, the cool gas mass increases with cluster mass, from approximately $10^{11} \,\rm M_\odot$ to $10^{12} \,\rm M_\odot$. The running average of cool gas locked in satellite galaxies within $2R_{200c}$ (including satellites of other FoF halos; dashed line) is everywhere only a factor of a few below the total running average (solid line), both for the cool gas mass fraction and total cool gas mass. This shows that a large fraction of cool gas is gravitationally bound in satellite galaxies at $z=0$. The individual data points are color-coded by the total instantaneous star formation rate of each cluster, including all cluster galaxies. At fixed halo mass, we identify a clear trend: clusters with higher cool gas content have higher star formation activity.

\begin{table*}[tbh]
    \caption{Selected cluster properties that we investigate for possible correlations with the cool gas mass fractions in clusters. Gas in the halo refers to within the FoF group, while gas in the central galaxy must also be gravitationally bound.}
    \label{tb:results:cluster_properties}      
    \centering          
    \begin{tabularx}{\textwidth}{lX} 
    \hline\hline
        Symbol & Explanation \\ \hline
        SFR & Instantaneous total star formation rate of the halo. \\
        $\rm SFR_c$ & Instantaneous total star formation rate of the central galaxy. \\
        $Z$ & Mass-weighted mean metallicity of all gas in the halo. \\
        $Z_{\rm c}$ & Mass-weighted mean metallicity of all gas in the central galaxy. \\
        $N_{\rm sat}$ & The number of satellites in the halo with stellar masses above $10^9 \,\rm M_\odot$. \\
        $M_{*,\rm halo}$ & Stellar mass of the halo. \\
        $M_{*,\rm central}$ & Stellar mass of the central galaxy, including all gravitationally-bound stars. \\
        $M_{\rm SMBH, tot}$ & Total mass of all SMBHs in the halo. \\
        $M_{\rm SMBH}$ & Mass of the most massive SMBH in the halo (typically in the central galaxy). \\
        $\dot{M}_{\rm SMBH, tot}$ & Total instantaneous mass accretion rate of all SMBHs in the halo. \\
        $\mu_{\rm mode}$ & A measure for the SMBH feedback mode of the most massive SMBH: the Eddington ratio normalized by the switchover threshold $\chi$ of the SMBH, i.e. $\mu_{\rm mode} = (\dot{M} / \dot{M}_{\rm Edd}) / \chi$. \citep[see][]{Weinberger2017TNG}. \\
        $E_{\rm cum, tot}$ & Total cumulative energy injected by the most massive SMBH of the cluster over its lifetime. \\
        $E_{\rm cum, kin}$ & Cumulative energy injected by the most massive SMBH while in the kinetic feedback mode. \\
        $E_{\rm cum, th}$ & Cumulative energy injected by the most massive SMBH while in the thermal (quasar) feedback mode. \\
        $E_{\rm kin} / E_{\rm tot}$ & Cumulative energy injected in kinetic mode by the most massive SMBH of the cluster, normalized by the total cumulative energy ejected irrespective of feedback mode. \\
        $M_{\rm kin} / M_{\rm tot}$ & Cumulative mass accretion of the most massive SMBH of the cluster while in kinetic mode over its total accreted mass. \\
        $M_{\rm cum, tot}$ & Total cumulative mass accreted by the most massive SMBH of the cluster over its lifetime. \\
        $M_{\rm cum, kin}$ & Cumulative mass accreted by the most massive SMBH while in the kinetic feedback mode. \\
        $M_{\rm cum, th}$ & Cumulative mass accreted by the most massive SMBH while in the thermal (quasar) feedback mode \\
        $N_{\rm SMBH, prog}$ & Number of progenitors of the most massive SMBH of the cluster, i.e. number of SMBH mergers. \\
        $\delta r$ & Relaxedness, defined as the distance between the center of mass and the position of the most bound particle, normalized by $R_{200c}$ \citep{Ayromlou2024AtlasTNGCluster}. \\
        $M_{\rm central} / M_{\rm tot}$ & Relaxedness, defined as the ratio between the mass of the central subhalo and the total \citep{Ayromlou2024AtlasTNGCluster}. \\
        $z_{\rm form}$ & Formation redshift of the cluster, i.e. when the halo obtained half of its mass $M_{200c}$ \citep{Ayromlou2024AtlasTNGCluster}. \\
        CCT & Central cooling time \citep[see][]{Lehle2024CoolCores}. \\
        \hline                  
    \end{tabularx}
\end{table*}

In general, we anticipate that the total cool gas content of a cluster could be correlated with many galaxy or halo properties. We therefore explore trends with 24 selected cluster properties, as described in detail in Table \ref{tb:results:cluster_properties}. The bottom panel of Figure~\ref{fig:results:cool_gas_vs_cluster_mass} shows the Pearson correlation coefficient $R$ between each property and halo cool gas fraction. For each property, we consider seven mass bins of 0.2 dex, depicted as colored dots. Crosses with error bars mark the mean correlation coefficient over the seven mass bins, and the standard deviation. We find correlation coefficients between $-0.5$ and $0.5$, indicating generally weak correlations. Indeed, most $R$ coefficients are consistent with zero across halo masses, or show non-monotonic halo mass trends indicative of low number statistics. Similar trends are present if we compute $R$ on semi-log or log-log scales.

However, several cluster properties exhibit a clear correlation with the cool gas fraction. These include various measures of star formation rate, stellar and SMBH masses, certain AGN feedback energetics, and aspects of the assembly history, such as the number of SMBH mergers and formation redshift. We also examine the correlation coefficients using cool gas mass instead of fraction (not shown). In most cases, these coefficients are consistent with zero, even for the aforementioned properties, indicating a lack of significant correlations with the absolute cool gas mass.

The only exception is the instantaneous star formation rate, for both the entire halo (SFR) and the central galaxy only ($\rm SFR_c$). Both exhibit positive correlations with both cool gas fraction and cool gas mass. The SFR shows the strongest correlation with the logarithm of the cool gas fraction and cool gas mass, with a mean correlation coefficient of $R \sim 0.5$. This link is physically expected as gas SFR depends on the existence, and amount, of dense (cool) gas. The global halo SFR sums over all satellite galaxies belonging to the FoF halo, reflecting a trend between halo cool gas and the number star-forming satellites (not shown). However, no significant trend exists with the total number of satellites with stellar mass $>10^9\, \rm{M_\odot}$, suggesting that perhaps only recent-infalling, gas-rich satellites correlate with the amount of cool cluster gas today \citep{Rohr2024CoolerPastClusters}. As we explore below, cool gas content in the cluster core is occasionally though not always present. The central galaxy $\rm SFR_c$ and global halo SFR differ substantially, but both correlate with the halo-scale cool gas fractions.

\begin{figure*}
    \centering
    \includegraphics[width=\textwidth]{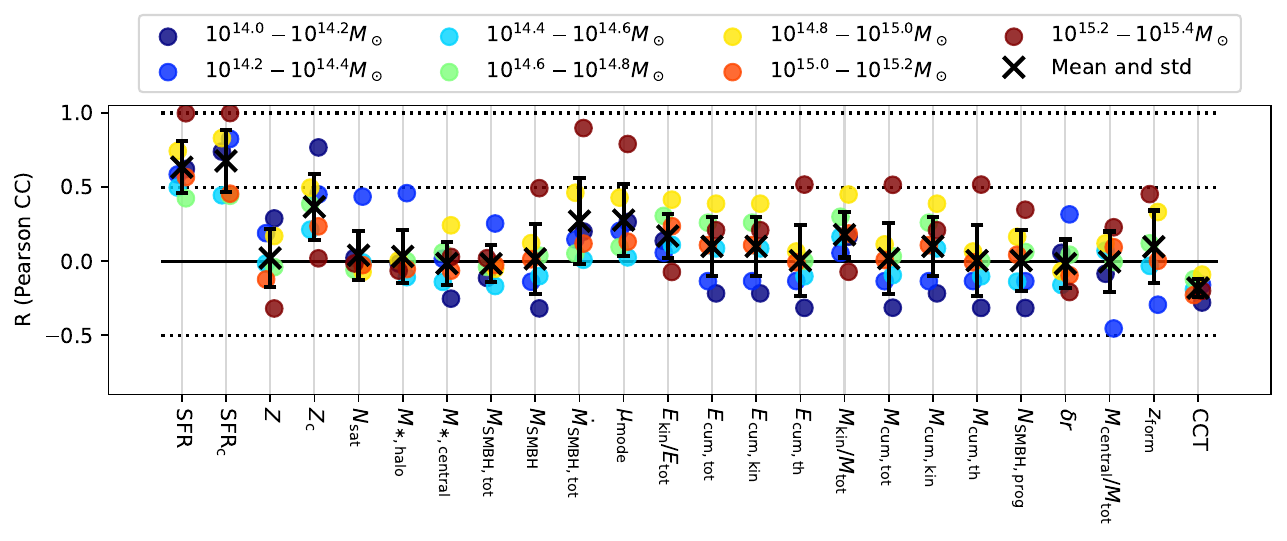}
    \caption{As in Figure~\ref{fig:results:cool_gas_vs_cluster_mass} but for the gas at the center of clusters. Namely, we show the Pearson correlation coefficients between cool gas fraction and halo and galaxy properties but here, instead of accounting for the cool gas within the entire $2 R_{200c}$ halo region, we consider it within the cluster core, $R < 0.05 R_{200c}$. The properties considered are described in Table \ref{tb:results:cluster_properties}. We split the sample into seven 0.2 dex mass bins (colored circles). The points have a small horizontal scatter for visual clarity. The black crosses mark the mean correlation coefficient over all mass bins, including error bars for standard deviation. Cool gas mass fraction within the central $5\%$ of the virial radius correlates strongly with star-formation rate, measured both in the central galaxy and integrated over all galaxies of the cluster.}
    \label{fig:results:cool_gas_vs_cluster_mass_core_only}
\end{figure*}

Figure~\ref{fig:results:cool_gas_vs_cluster_mass_core_only} repeats this analysis for cluster cores, showing the same correlation coefficients for cool gas within $0.05 R_{200c}$. In general, trends are similar and similarly weak, with several notable exceptions. The correlation between star formation rate -- both across the entire halo and the central galaxy -- is strong, with correlation coefficients of $\sim 0.6 - 0.7$. In addition, the metallicity of the central galaxy $Z_c$ and the central cooling time (CCT) show weak correlations with cool gas. Central metallicity $Z_c$ is positively correlated with cool gas fraction and mass with $R \sim 0.4$ for both, underlining the link between the existence of cool gas in the central regions with star formation, which enriches gas in the central regions of the cluster. The central cooling time is negatively correlated with both cool gas fraction and cool gas mass with $R \sim 0.2$, indicating that longer cooling times in the cores of clusters are related to a lower cool gas content in the inner regions \citep{Lehle2024CoolCores}. Some SMBH properties also show weak trends with cool gas abundance, indicating that SMBH feedback is more important for cool gas in the central regions of clusters compared to their outskirts.

\subsection{Spatial distribution of cool gas in clusters}
\label{ssc:results:radial_distribution}

\begin{figure*}
    \centering
    \includegraphics[width=\textwidth]{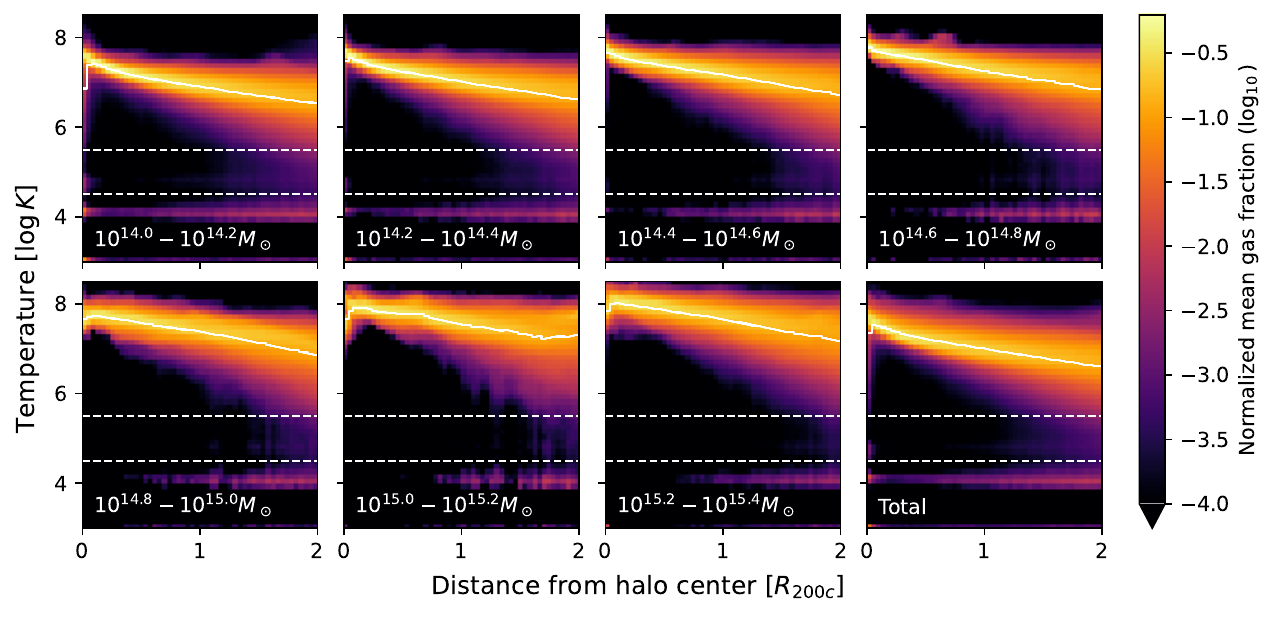}
    \caption{Mean radial temperature distribution of all clusters in TNG300 and TNG-Cluster at $z=0$. The color of each pixel shows the mean gas mass at that temperature and distance, averaging over all clusters in seven mass bins of 0.2 dex (first seven panels, bin range denoted in bottom left of every panel), plus the total mean profile over all clusters (final panel). Each panel is then normalized, giving fractions. The white solid lines show the running temperature average, the white dashed lines show the division between temperature regimes. The profiles are normalized such that in each distance bin, the values across all temperature bins sums to unity. Cool gas is predominantly located in the outskirts and the center of clusters.}
    \label{fig:results:radial_dependence_temperature}
\end{figure*}

Figure~\ref{fig:results:radial_dependence_temperature} presents the temperature distributions of all gas within $2R_{200c}$ for the full cluster sample at $z=0$. The first seven panels display the marginalized temperature distributions at different distances, binned by halo mass. The final panel combines all halos. The profiles are normalized such that, at a fixed radius, the color values sum to unity, with the mean taken across the halos in each panel. The white solid lines represent a running mean radial temperature profile, while the white dashed lines indicate the boundaries of the three temperature regimes at $10^{4.5} \,\rm K$ and $10^{5.5} \,\rm K$.

The characteristic features can be seen in the final panel (bottom right), where we identify the hot phase at $\sim 10^8 \,\rm K$ in the cluster centers, dropping to $\sim 10^7 \,\rm K$ by the virial radius. We also see a transitional warm phase at $\sim 10^5\,\rm K$, a cool feature at $\sim 10^4\,\rm K$, and a stripe of star-forming gas, placed at $10^3\,\rm K$. At all distances, hot gas dominates by mass, forming a broad distribution in temperature. The mean radial profiles fall within the hot phase everywhere.

The width of this hot phase temperature distribution increases with distance, to the point that it spans roughly two orders of magnitude in temperature at the virial radius, and becomes even wider by $2R_{200c}$. The broad temperature distribution of the hot halo gas is not due to stacking -- it is also present for individual clusters (not shown). In the centers of clusters, the average temperature peak typically exceeds the virial temperature by a factor of a few.

The warm band at $\sim 10^5\,\rm K$ is faint, only being visible in some cluster centers and beyond the virial radius, perhaps where the ICM meets the IGM. The cool gas fraction is maximal (i) in the central core of the cluster, and/or (ii) in the cluster outskirts. On average, however, clusters contain little cool gas in their cores \citep{Lehle2024CoolCores}, and this peak in cool gas fraction is the effect of a few clusters with high cool gas content at their centers. Finally, the abundance of star-forming gas has a similarly strong contribution at the cluster center, in addition to an increasing presence towards cluster outskirts. At large distances, there is a continuous distribution of gas temperatures and no clear separation between the cool, warm, and hot phases.

Across the first seven panels, several mass trends are apparent. The band of hot gas starts at lower temperatures in the lower mass bins, due to the lower virial temperature of lower mass clusters. Warm gas is present further into the cluster at lower cluster masses, with the lowest mass bin showing the warm gas still contributing gas fractions of $> 10^{-4}$ around the virial radius. In the highest mass bin at large distances, a fraction of warm gas exists. The cool gas exhibits similar behavior: at low cluster masses, it contributes a notable fraction to the overall gas at all distances, especially at the cluster center. With increasing cluster mass however, the cool gas becomes less significant within the virial radius. The star-forming gas behaves similarly, except at the cluster center, where at high cluster masses, star-forming gas makes up a small but visible fraction of the total gas content. The profiles of individual clusters show that this is a stacking effect caused by a few massive clusters having star-forming gas at their centers, as opposed to all halos at these masses having small amounts of star-forming gas (not shown). We also observe that cool and star-forming gas throughout the halo are patchy, as indicated by their clumpy radial distributions \citep[see also][]{Rohr2024CoolerPastClusters}. This hints that individual substructures, i.e. satellites, may host a large fraction of this cool gas, rather than it being smoothly distributed throughout the ICM. 

\begin{figure*}
    \centering
    \includegraphics[width=\textwidth]{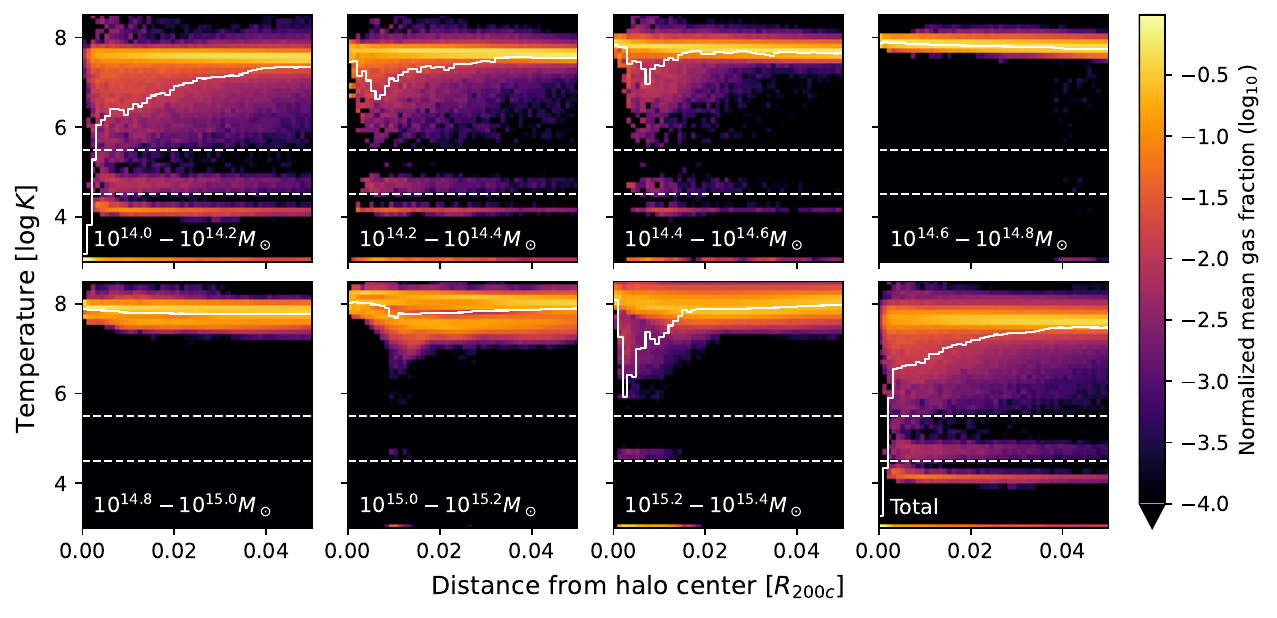}
    \caption{Same as Figure~\ref{fig:results:radial_dependence_temperature} but limited to the central regions up to $0.05 R_{200c}$. Cool gas in cluster cores is found only rarely, primarily in the lowest as well as highest mass halos shown.}
    \label{fig:results:radial_dependence_temperature_core}
\end{figure*}

The central regions of the clusters exhibit a unique temperature structure. Figure~\ref{fig:results:radial_dependence_temperature_core} shows similar marginalized temperature distributions as a function of radius, but limited to the inner 5\% of the virial radius. At all masses, the hot phase distribution again dominates. However, towards the cluster core substantial fractions of this gas transitions into the warm phase \citep[see][]{Reefe2025}, and the running average (white line) drops to lower temperatures. Cool and warm gas phases are abundant at $10^4$ and $10^5\,\rm K$ respectively, with the warm gas being the smaller by mass contribution. Star-forming gas is particularly abundant for low-mass clusters and even dominates the innermost radial bin(s), leading to the visible suppression of all other temperatures at the smallest radii.

The temperature profiles of the inner regions vary significantly across different mass bins. At the low-mass end, the hot gas exhibits a broader distribution, and the warm and cool gas extend out to approximately 5\% of the virial radius. As cluster mass increases, the distribution of hot gas temperatures at each radius becomes progressively narrower—initially at larger distances and later in the core—until, in the mass range $10^{14.6 - 15.0} \,\rm M_\odot$, the hot gas is confined to a narrow band around $10^8 \,\rm K$. However, for even more massive clusters, this trend appears to reverse, with the hot gas temperature distribution broadening again at smaller radii, accompanied by the reappearance of warm-phase cooling channels.

The fractions of cool and warm gas decrease with increasing cluster mass, becoming nearly absent in the intermediate mass bins (top right and bottom left panels). Interestingly, only in the highest mass bin (second to last panel) do the warm and cool gas fraction start to reappear in noticeable amounts.

In general, the warm, cool, and star-forming gas visible in these stacked analyses is contributed by a small fraction of clusters. The individual profiles for most clusters show exclusively hot gas within 5\% of the virial radius. We return below to this unique sub-population of galaxy clusters with significant amounts of warm, cool, and star-forming gas in their cores.

\begin{figure*}
  \centering
  \includegraphics[width=0.7\textwidth]{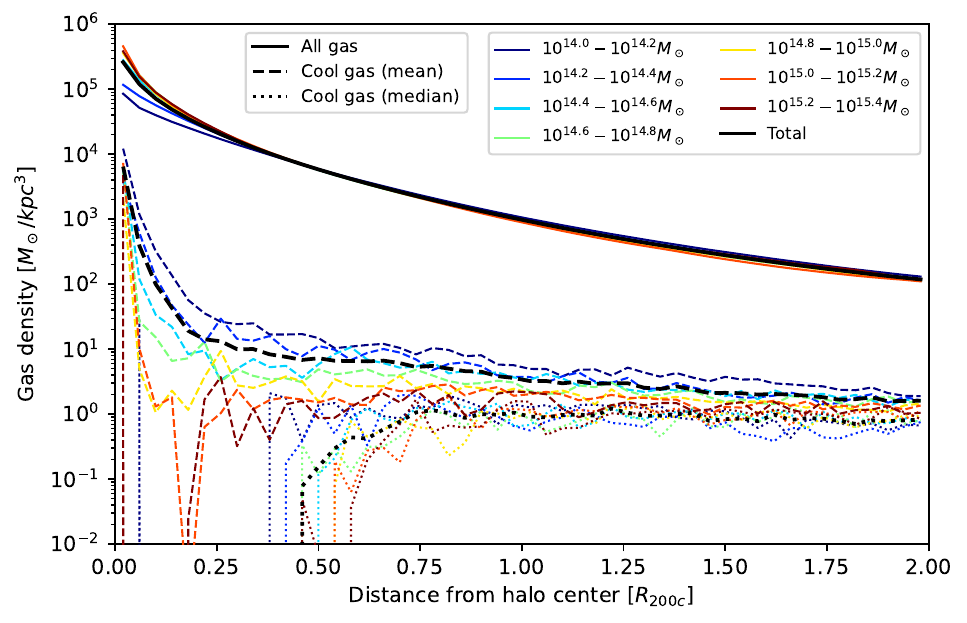}
  \includegraphics[width=0.48\textwidth]{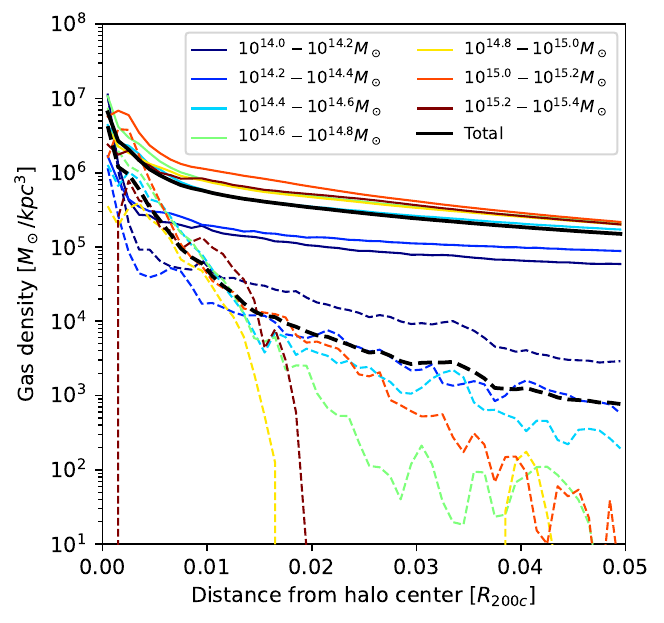}
  \includegraphics[width=0.48\textwidth]{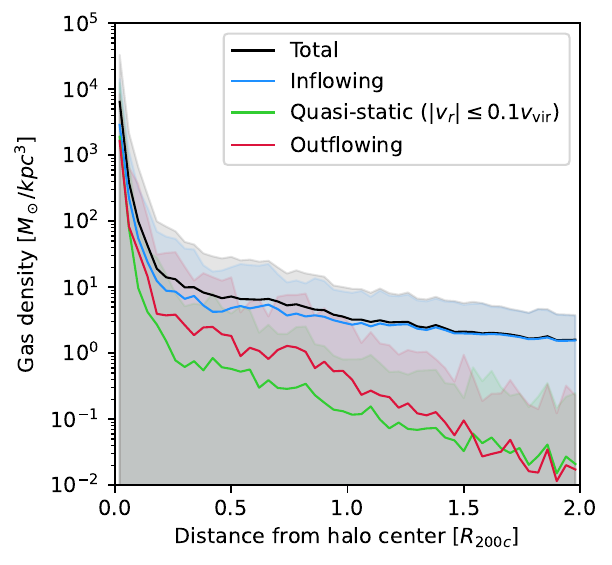}
  \caption{Gas density profiles of clusters in TNG300 and TNG-Cluster at $z=0$ for gas in different temperature phases and with different radial velocities. Top: Gas density profile out to $2R_{200c}$. The solid lines show the mean total gas density whereas the dashed (dotted) lines show the mean (median) density of only the cool gas ($T < 10^{4.5} K$). The black line shows the mean across the entire sample of clusters, while the colored lines show the mean in mass bins of 0.2 dex. Bottom left: Same as top, but only out to $0.05R_{200c}$. Bottom right: Density profile of cool gas for all clusters, split by the radial velocity of the gas. Lines show the mean density profile for cool gas of all clusters for all cool gas (black line), inflowing cool gas (blue), quasi-static cool gas (green) and outflowing cool gas (red). Gas is considered inflowing (outflowing) if its radial velocity is towards (away from) the cluster center, and quasi-static if its absolute radial velocity is below 10\% of the cluster virial velocity. The shaded regions show the standard deviation. Cool gas has lower average density than the overall cluster gas and is predominantly inflowing into the cluster, except in the cluster center.}
  \label{fig:results:radial_density_profiles}
\end{figure*}

To quantify the spatial distribution of cool gas in clusters, Figure~\ref{fig:results:radial_density_profiles} shows radial density profiles of all gas (solid lines) versus cool gas (dashed lines). We also contrast against the median density profiles of cool gas (dotted lines). Each is given for the full sample (black) and in 0.2 dex mass bins (colored lines). The total gas density rises steadily from $2R_{200c}$ towards the cluster center. In the halo outskirts, it shows remarkable invariance as a function of mass: clusters of higher masses have slightly lower densities beyond the virial radius. At the center, within $\sim 0.3R_{200c}$, the overall density of gas in lower mass clusters is lower than for higher mass clusters. 

The mean cool gas densities (dashed lines) on the other hand show a clear mass trend beyond $0.5 R_{200c}$, with higher mass clusters having lower cool gas densities. The cool gas density profiles in this radial range are also relatively flat, dropping only by a factor of $\sim5$ over a distance of 1.5 virial radii. Within $0.5 R_{200c}$, the mean cool gas density rises sharply, crossing three orders of magnitude towards the center. While the mass trend holds, the more massive clusters experience a drop in mean cool gas density between the core of the cluster and $0.25 R_{200c}$, with the magnitude of this drop increasing with cluster mass.

The median cool gas density profiles show no discernible mass trend outside of $0.75 R_{200c}$, lying below their corresponding mean values. Inside of $0.75 R_{200c}$ the median density rapidly drops off to zero. The radius where this drop-off occurs has a clear mass trend, with more massive clusters having zero cool gas in the median at larger distances. This means that for at least half of all clusters, the gas density in the inner $\sim 50\%$ of the clusters virial radius is below the resolution limit of the simulations $\lesssim 10^{7}\,\rm{M_\odot}$. Most clusters simply contain little to no cool gas in this region. That is, cool gas tends to avoid the inner regions of massive clusters, with more massive clusters exhibiting larger regions in their interior devoid of cool gas. This is consistent with findings of \cite{Kleiner2023MeerKAT}, who find that \metalline{H}{i}-detected dwarfs avoid the central regions of the Fornax cluster, and those of \cite{Burchett2018WarmHot}, who find the atmospheres of cluster galaxies increasingly \metalline{H}{i}-depleted towards the denser cluster central regions. 

A notable exception to this trend is the cluster centers, i.e., within the central galaxies. In particular, the mean cool gas density rises strongly towards the center of clusters, even for the most massive clusters that have a cool gas deficiency at intermediate distances. The bottom left panel of Figure~\ref{fig:results:radial_density_profiles} shows the density profile inside of $0.05 R_{200c}$. Strikingly, the densities of cool gas reach values comparable to the total gas density at the center of clusters. The same mass trends hold, with more massive clusters showing higher overall gas densities. However, cool gas in the core has no clear trend with halo mass. The most massive clusters show a drop in cool gas density beyond a few per-cent of the virial radius, while the lower mass clusters only show a gradual decrease in their mean cool gas density, but the position of the drop-off follows no clear mass trend. The median density is zero everywhere in the central region and therefore not depicted. This again emphasizes that the interior of most clusters is dominated by hot gas, where a small subset have substantial reservoirs of cool and warm gas content within 5\% of their virial radius, and these dominate the mean density profiles.

\subsection{Velocity distribution of cool gas in clusters}
\label{ssc:results:velocity_distribution}

As first step towards investigating the origin of the cool gas, Figure~\ref{fig:results:radial_density_profiles} examines the radial velocity of the cool gas (bottom right panel). We show the cool gas density profile, decomposing gas by its radial velocity. The black line shows the total mean cool gas density. The three colored lines show only cool gas that is moving into the cluster (`inflowing', blue), cool gas moving out of the cluster (`outflowing', red) and quasi-static cool gas, defined as radial velocities smaller than 10\% of the virial velocity\footnote{The results are qualitatively the same if we instead choose a threshold in unscaled km/s with a value similar in order of magnitude to 10\% of the mean virial velocity of the clusters, i.e. $\sim 100\,\rm km/s$.} of their cluster (`quasi-static', green). The curves show mean profiles across the cluster sample, and the shaded regions give the standard deviations.

Inflowing cool gas overwhelmingly dominates over outflowing cool gas almost everywhere from $0.2 R_{200c}$ out to the halo outskirts. Nevertheless, within the central region ($R<0.2 R_{200c}$), outflowing and quasi-static components contribute at comparable levels to the inflowing gas, though the latter still remains dominant. In the outskirts, beyond the virial radius, the gas density is almost entirely dominated by gas moving toward the cluster center.\footnote{We do not include the Hubble expansion correction in these radial velocity calculations.} This supports the previous findings of \cite{Ayromlou2024AtlasTNGCluster}, which argued that cool gas is predominantly inflowing, whereas most of the outflowing gas is hot.

\begin{figure*}
    \centering
    \includegraphics[width=0.24\textwidth]{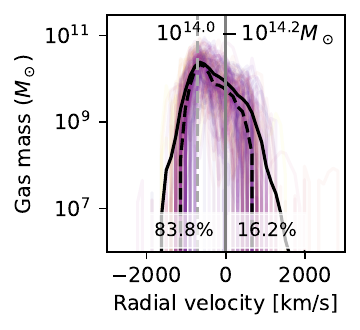}
    \includegraphics[width=0.24\textwidth]{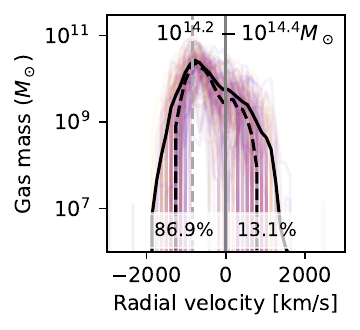}
    \includegraphics[width=0.24\textwidth]{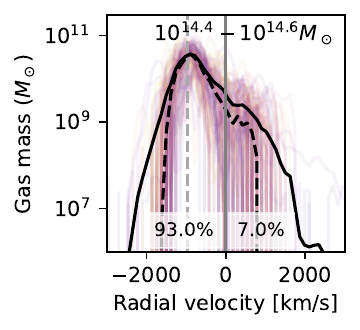}
    \includegraphics[width=0.24\textwidth]{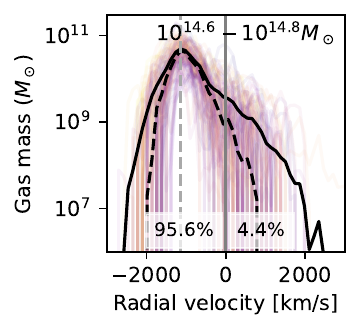}
    \includegraphics[width=0.24\textwidth]{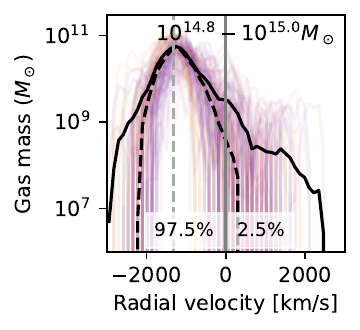}
    \includegraphics[width=0.24\textwidth]{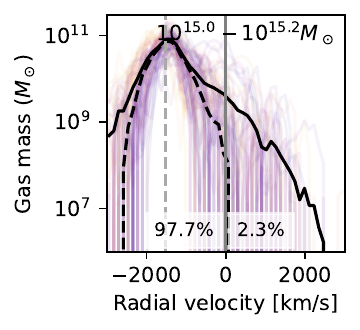}
    \includegraphics[width=0.24\textwidth]{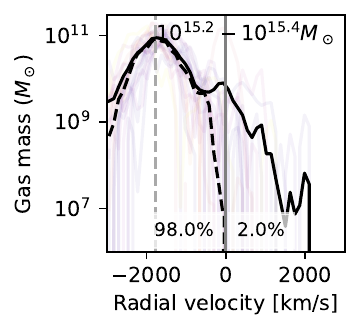}
    \includegraphics[width=0.24\textwidth]{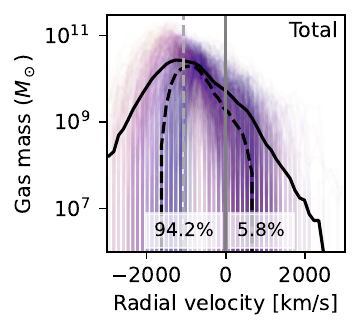}
    \includegraphics[width=0.5\textwidth]{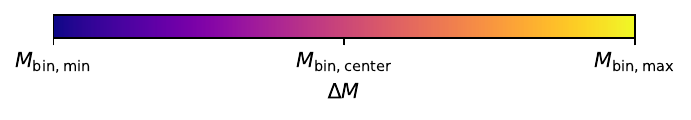}
    \caption{Radial velocity distribution of cool gas in clusters from TNG300 and TNG-Cluster at $z=0$. Each panel shows the velocity distribution of cool gas cells for every cluster weighted by the mass of each gas cell, in seven cluster mass bins of 0.2 dex. The faint colored lines show the velocity distribution of individual clusters and the solid (dashed) black line shows the mean (median) of the mass bin. The last panel shows the distribution of all clusters. Negative velocity denotes gas moving towards the cluster center (inflowing), positive velocity denotes gas moving away from the cluster center (outflowing). The individual distribution lines are colored according to the host cluster mass: An orange color denotes a mass near the high edge of the mass bin, a purple color a mass near the low edge of the mass bin. The grey solid line marks the position of $v_r = 0$; the label to either side denote the mean fraction of gas mass to the left (negative velocity, infalling) and right (positive velocity, outflowing) of the zero line. The grey dashed line marks the position of the negative of the mean virial velocity in the mass bin. The cool gas in clusters is predominantly infalling i.e. inflowing towards the center.}
    \label{fig:results:velocity_distribution}
\end{figure*}

Figure~\ref{fig:results:velocity_distribution} shows the distributions of radial velocities of cool gas in clusters. The first seven panels show the same 0.2 dex mass bins as above, while the final panel combines all clusters. The colored lines show the distribution for individual halos: color denotes mass within the respective mass bin. The solid (dashed) black line shows the mean (median) velocity distribution over all clusters in each panel. The percentages at the bottom show the mass fraction of inflowing vs. outflowing gas. 

We see that the majority of cool gas is inflowing with radial velocities towards the cluster center. This shows that even today at $z=0$, clusters still accrete cool gas. The peak of the mean and median velocity distributions is always close to the virial velocity (vertical gray dashed lines). This implies that cool gas falls into the cluster at this characteristic speed, i.e. with motion due to gravity. Secondary, small peaks are visible around zero radial velocity, reflecting quasi-static cluster ICM. The density profiles above suggest that most of this quasi-static mass is preferentially found at small distances, well within the halo. More massive clusters have a larger mass fraction of inflowing cool gas, due to their deeper gravitational potentials. 

\subsection{The origin of cool cluster gas: assembly history}
\label{ssc:results:the_origin_of_cool_cluster_gas}

We turn to the origin of this $z=0$ cool gas in and around clusters. To do so, we use the Monte Carlo tracer particles contained within all cool gas cells within $2 R_{200c}$ (as described in Section~\ref{ssc:methods:analysis_choices}). We focus on the 352 halos of TNG-Cluster, and follow these tracers back in time until redshift $z = 8$. At progressively earlier epochs, the gas associated with these tracers may be in different phases, other halos, or unbound altogether and located in the IGM.

\begin{figure*}
    \centering
    \includegraphics[width=0.52\textwidth]{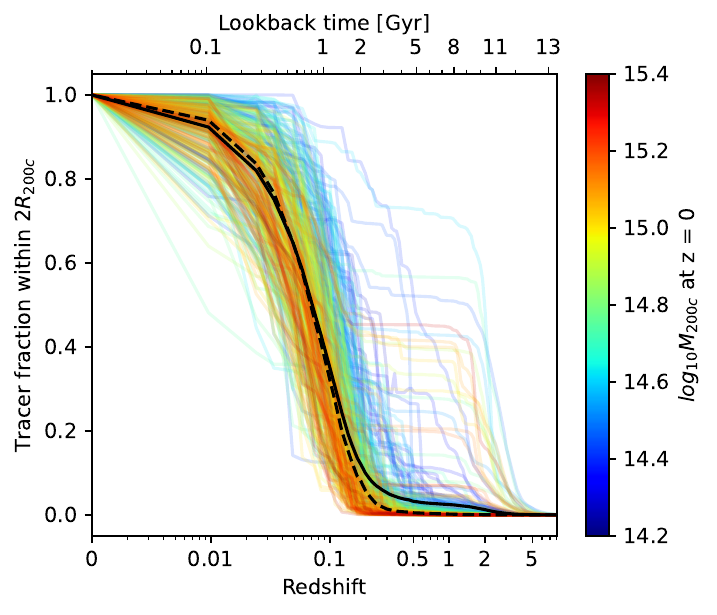}
    \includegraphics[width=0.43\textwidth]{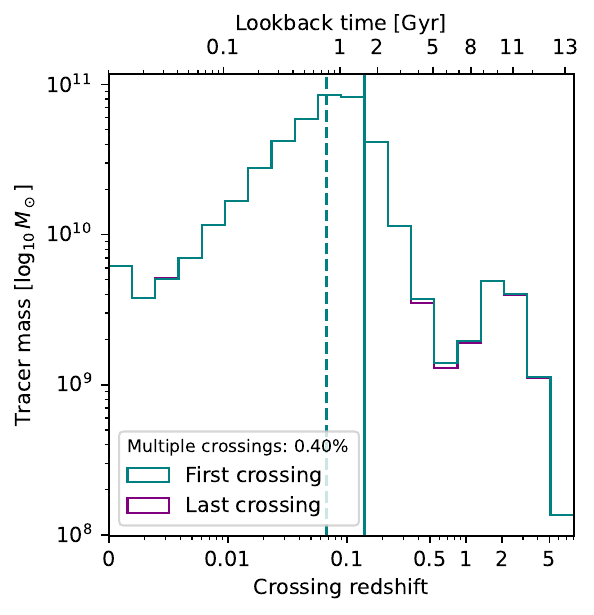}
    \includegraphics[width=0.48\textwidth]{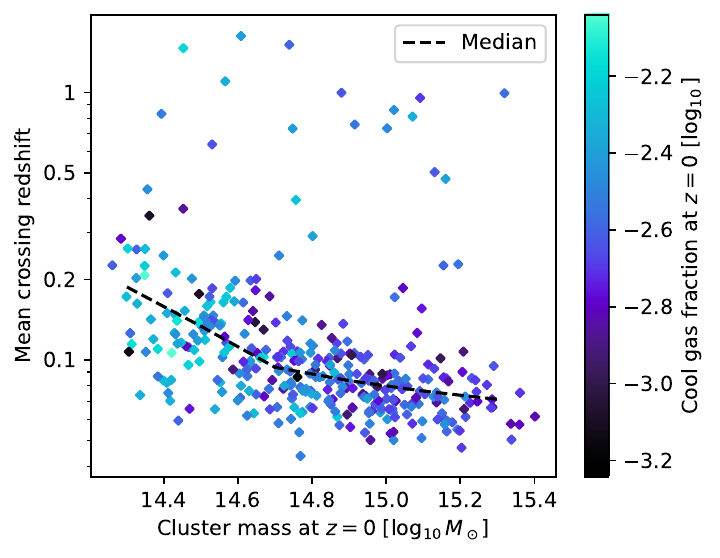}
    \includegraphics[width=0.48\textwidth]{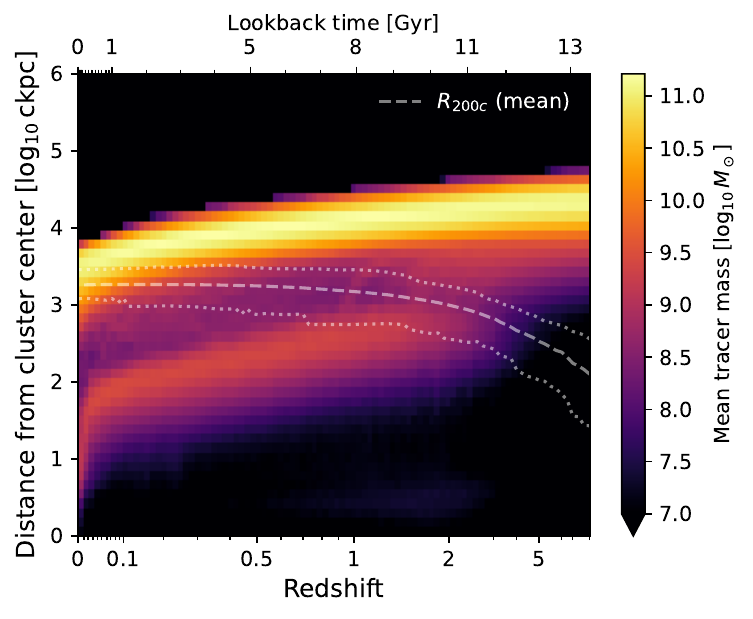}
    \caption{Evolution of the spatial distribution of tracers that contribute to $z=0$ cool gas in clusters, from TNG-Cluster. In practice, here we follow those tracers that at $z=0$ are associated with gas cells of temperatures below $10^{4.5}\,\rm K$ and within $2R_{200c}$ of their host cluster. Top left: Fraction of selected tracers within $2R_{200c}$ as a function of redshift. The virial radius $R_{200c}$ develops with redshift and is that of the cluster progenitor which eventually hosts the cool gas at redshift $z = 0$. Top right: Mean distribution of crossing redshifts, defined as the first (green) or last (purple) redshift at which the distance from a tracer to the progenitor of its $z = 0$ host cluster decreases below $2R_{200c}$. The solid (dashed) vertical line shows the mean (median) crossing redshift. Bottom left: Mean $2R_{200c}$ crossing redshift of every cluster in TNG-Cluster versus the $z = 0$ mass of the cluster. The individual points are colored by the cluster cool gas mass fraction at $z = 0$, while the black dashed line shows the median trend. Bottom right: Marginalized distribution of distances of tracers as a function of redshift. Color indicates the bin-wise mean tracer mass, per bin per cluster. The dashed white line shows the mean virial radius of the cluster sample, and the two dashed white lines show the minimum and maximum virial radius of the sample in each redshift bin. Clusters gather most of the gas that eventually forms redshift $z=0$ cool gas around $z \sim 0.1$.}
    \label{fig:results:tracers_dist}
\end{figure*}

Figure~\ref{fig:results:tracers_dist} shows the fraction of selected tracers that are within a sphere of radius $2R_{200c}$, as a function of redshift (upper left). To do so, we follow the evolving virial radius and center position of each cluster progenitor with time. The colored lines show individual clusters, with color denoting halo mass at $z = 0$, while the solid (dashed) black line shows the mean (median) over all clusters. 

From early times towards $z=0$, the tracks of individual clusters tend to remain low (<10~per-cent) at $z \gtrsim 0.2$ until $z \sim 0.1$, where they then increase sharply to values of unity. This shows that the majority of cool gas enters the $2R_{200c}$ region at similar, and late, times. We see that scatter with respect to this typical evolution is set by $z=0$ halo mass: more massive clusters (red) assemble their cool gas later, while lower mass clusters (blue) assembly their cool gas earlier. This is consistent with formation redshift decreasing as a function of increasing halo mass under hierarchical structure assembly. However, recall that we show the fraction of tracers within $2R_{200c}$ having selected specifically those tracers within $z=0$ cool gas. The remarkably rapid rise of this fraction at a characteristic $z \sim 0.1$ therefore also shows that either (i) cool gas that entered the cluster progenitors at earlier times is a small fraction of its final cool halo gas mass, and/or that (ii) cool gas cannot remain cold, such that the $z=0$ cool gas preferentially traces recent accretion.

For every tracer, we identify the redshift when it crosses a distance of $2R_{200c}$ of its host cluster progenitor from the outside. A tracer can in principle leave the sphere of radius $2R_{200c}$ again after having entered it once, due to feedback-driven outflows or bulk motions. We therefore consider the redshift of the first and the last crossing separately. For tracers that only cross $2R_{200c}$ once, these are identical.

Figure~\ref{fig:results:tracers_dist} shows the distribution of the first and last crossing redshifts of $2R_{200c}$ in green and purple, respectively (upper right). The mean and medians of both are shown as solid and dashed lines, respectively. Both distributions are nearly identical: only 0.4\% of tracers cross the $2R_{200c}$ sphere more than once, indicating that the vast majority of gas falls into a cluster once and never leaves. The mean (median) crossing redshift is $z = 0.14$ ($z = 0.07$). This again shows that clusters assemble most of their cool gas only relatively late. There is a secondary, smaller peak in the distribution, around $z \sim 2$. This corresponds to a subset of clusters that gather a large fraction of their cool gas at earlier times. Due to the small difference between the first and last crossing redshifts, we adopt the more general term `crossing redshift' hereafter, always referring to the first $2R_{200c}$ crossing redshift of each tracer.

Quantifying the halo mass dependence, Figure~\ref{fig:results:tracers_dist} shows the mean crossing redshift for all 352 clusters in TNG-Cluster as a function of $z=0$ halo mass (lower left). The color of each marker gives cool gas fraction within $2R_{200c}$ at $z = 0$. The majority of halos have low mean crossing redshifts around $z \sim 0.1$, and for these a downward trend with cluster mass exists, as seen above and indicated here with the dashed black median line. The scatter is large, suggesting that other factors besides the cluster mass play an important role in the assembly history of cool gas. There are a handful of outliers with mean crossing redshifts $z > 0.5$, and we speculate these outliers could be driven by specific major merger events. No discernible trend with $z = 0$ cool gas fraction at fixed cluster mass is apparent.

Finally, to investigate the different trajectories in distance that accreting gas can take, Figure~\ref{fig:results:tracers_dist} shows the evolving distributions of tracer distances (lower right). We show the bin-wise mean distribution over the entire cluster sample, with the color giving the mean tracer mass per bin and per cluster. The white dashed line shows the mean virial radius of the sample, while the faint dotted lines bracket the minimum and maximum. Most of the cool gas follows a well-defined track that begins at $\sim 10^{4-4.5}\,\rm ckpc$ and gradually decreases until reaching the $z=0$ distance distribution. Most of the tracer mass is in this primary track at all redshifts, making it the most important channel. It is beyond the virial radii until $z < 0.1$, as expected since only $\sim$\,20\% of the cool gas in our selection reaches distances within the virial radius.

A second, lower track is also visible beginning at redshift $z \sim 4$, at a much smaller distance of $\sim 10^{2.5}\,\rm ckpc$, just below the mean virial radius of the cluster sample. This indicates that the material had already been incorporated into the ICM of the cluster progenitors by this time. It follows the same slope as the primary track but extends to a much lower final distance of $10 - 100\,\rm ckpc$ at redshift zero. Between these two tracks, individual curved lines are visible, marking signatures of mergers that coherently bring in large amounts of material. Additionally, a faint but detectable region appears at distances near zero between $z=1$ and $z=2$, which then fades at lower redshifts. This corresponds to cool gas in the central galaxy, which is eventually eliminated as these systems evolve into ultra-massive, quenched brightest cluster galaxies (BCGs).

The distance distributions for individual clusters reveal significant diversity (not explicitly shown). Some halos, particularly at the high-mass end, are predominantly characterized by the primary track described above, while others exhibit a more balanced distribution of mass between the distant and closer tracks. The trajectories of individual tracers are similarly varied, reflecting not only accretion from the IGM but also V-shaped profiles driven by several effects—such as backsplash galaxies carrying substantial amounts of cool gas, which enter the cluster at high velocities.

\begin{figure*}
    \centering
    \includegraphics[width=0.43\textwidth]{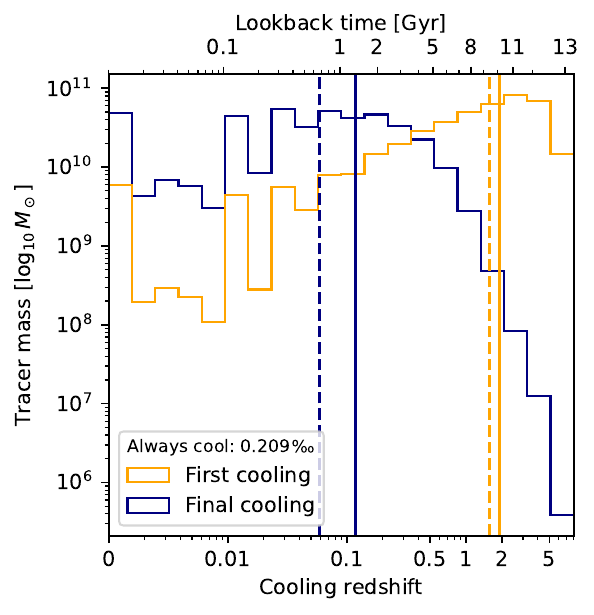}
    \includegraphics[width=0.53\textwidth]{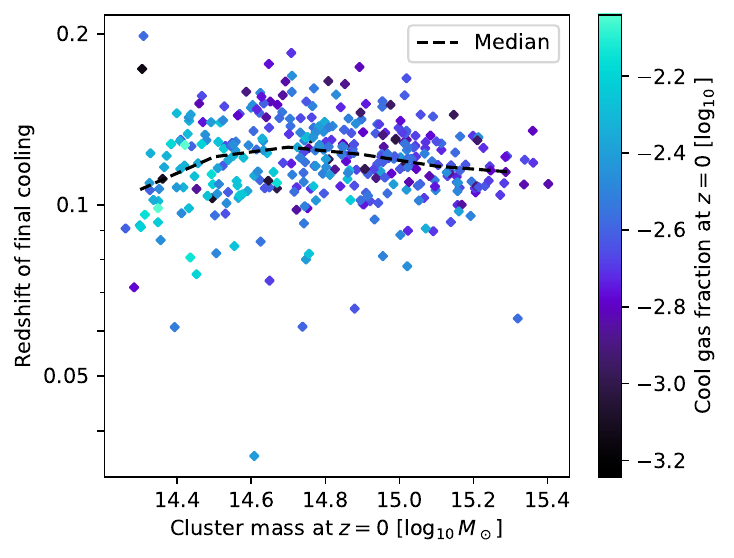}
    \caption{When did the $z=0$ cool gas in clusters cool down? As in Figure~\ref{fig:results:tracers_dist}, we use TNG-Cluster and the tracers that contribute to $z=0$ cool gas. Left: Mean distribution of cooling redshifts, defined as when gas associated with the selected tracers cools below the threshold value of $10^{4.5}\,\rm K$ for the first (orange) or last (blue) time. The solid (dashed) line marks the position of the mean (median) cooling redshift for each of the distributions. Right: mean redshift of final cooling per cluster as a function of halo mass at $z = 0$. Color denotes the cool gas fraction at redshift $z = 0$, while the black dashed line shows the median relation. The gas that eventually forms redshift $z=0$ cool gas typically cools and heats up multiple times before cooling for the final time around $z \sim 0.1$, potentially a signature of pre-processing other halos.}
    \label{fig:results:cooling_times}
\end{figure*}

\subsection{Temperature evolution and cooling}
\label{ssc:results:temperature_development}

Is the cool cluster gas at $z=0$ already cool at earlier times, or did it cool only recently? To address this question, we assign each tracer two cooling redshifts, representing the redshifts at which the associated gas first and last cools from high temperatures into the cool regime. \footnote{To account for tracers associated with stars, we assume the temperature of star particles to be in the warm or hot regime.}

Figure~\ref{fig:results:cooling_times} shows the distribution of these cooling redshifts for all clusters of TNG-Cluster (left panel). The first cooling times are shown in orange, and the last cooling times in blue. Solid and dashed lines mark the mean and medians of each distribution, respectively. As opposed to the crossing redshifts, the two distributions have markedly different shapes. The distribution for the first cooling peaks at higher redshifts, with a mean (median) of $z = 1.9$ ($z = 1.6$), with a tail towards lower redshift. In contrast, the distribution for the final cooling redshift is peaked strongly at low redshift, with a mean (median) at $z = 0.1$ ($z = 0.06$).

A negligible fraction ($\sim0.2$~per~cent) of gas has been cool since $z = 8$ until today, indicating that nearly all cool gas today has been heated above the background IGM temperature before cooling again, i.e. almost no gas survives infall into a cluster while remaining cool. The difference in the distribution of first and final cooling redshift also suggests that gas can cool and be reheated during its infall into the cluster, before finally cooling at lower redshift to form the redshift zero cool gas component of the cluster. Alternatively, pre-processing in lower mass halos, virial reheating during accretion into the main cluster progenitor, and eventual cooling $\sim 1-2$\,Gyr prior to $z=0$ could produce a similar signature. The mean and median redshift of the final cooling ($z\sim 0.1$) roughly coincide with the mean and median crossing redshift for cool gas predecessors crossing the $2R_{200c}$ sphere of their host cluster, meaning that gas typically cools down to cool temperatures for the last time as it enters the clusters' sphere within twice the virial radius, where it remains cool until today. A caveat to this is the tail of the distribution for final cooling towards low redshift, which shows a significant fraction of the redshift zero cool gas has only cooled to these temperatures recently. This tail for recently cooled gas also exists for gas that is cooling for the first time.

The right panel of Figure~\ref{fig:results:cooling_times} shows the mean redshift of final cooling for every cluster in the TNG-Cluster sample and its dependence on $z=0$ halo mass. The individual points are colored by the cool gas fraction of the cluster within $2R_{200c}$ at $z = 0$. The black dashed line shows the median relation with halo mass, which is nearly flat, albeit with large scatter. At fixed halo mass, there is no clear trend between when gas cools, and the amount of cool gas that exists at $z=0$.

\begin{figure*}
    \centering
    \includegraphics[width=0.49\textwidth]{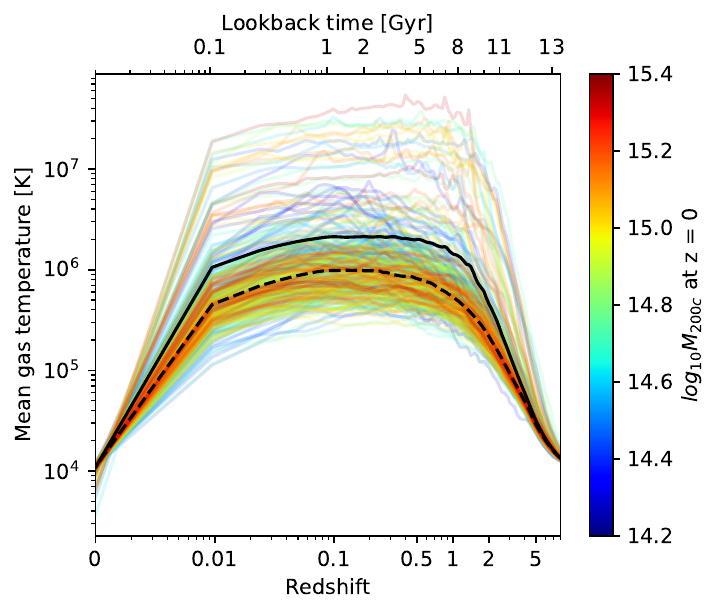}
    \includegraphics[width=0.49\textwidth]{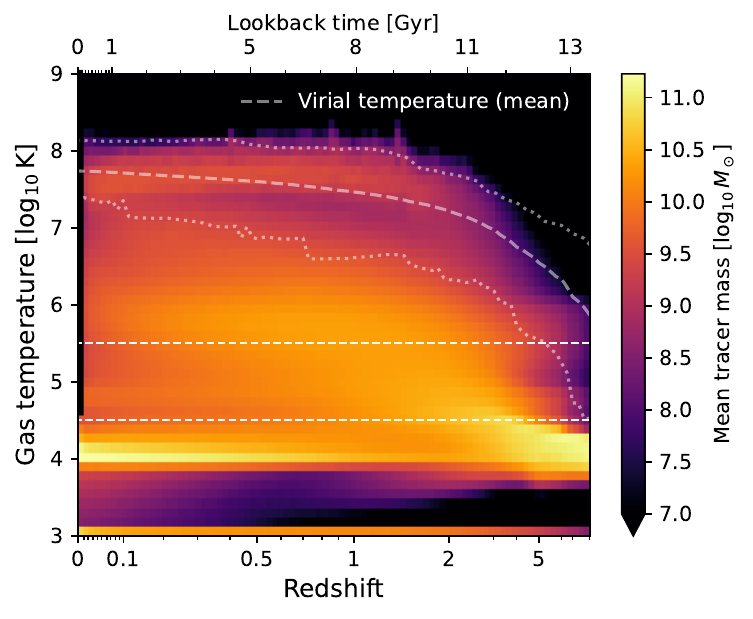}
    \caption{Evolution of the temperature of tracers that contribute to $z=0$ cool gas in clusters, from TNG-Cluster. Left panel: mean temperature of gas cells associated with the cool gas tracers in each cluster as a function of redshift. Each halo of TNG-Cluster is shown as a single line, with color denoting the $z=0$ cluster mass $M_{200c}$. The black solid (dashed) line shows the mean (median) over the entire cluster sample. Right panel: distribution of temperature for gas cells associated with the selected tracers with redshift, showing the bin-wise mean over the individual distributions of all clusters. The faint dashed white line shows the mean virial temperature of the cluster sample, and the two faint dashed white lines show the minimum and maximum virial radius of the sample in each redshift bin. The two straight dashed white lines show the division between the temperature regimes at $10^{4.5}$ (below which we talk of cool gas) and $10^{5.5}\,\rm K$ (above which we talk of hot gas), respectively. The predecessor gas to redshift $z=0$ cool gas on average spends most of the time since $z = 8$ at temperatures $> 10^5 \,\rm K$, but below the virial temperature of the host cluster, and cooled only recently.}
    \label{fig:results:temperature_evo}
\end{figure*}

Figure~\ref{fig:results:temperature_evo} shows the evolution of temperature as a function of redshift. The left panel shows the mean temperature of all gas cells associated with the selected tracers in each cluster colored lines, where the color denotes the cluster mass $M_{200c}$ at redshift zero. The black solid and dashed lines show the mean and median over all clusters, respectively. The general behavior is consistent across halos. First, gas starts to cool near $\sim 10^4$\,K at high redshift $z>6$, before the epoch of reionization and the UV background heats up the IGM. Between redshifts $z\sim2-6$, the average gas temperature increases to $\sim10^6$~K as a result of the onset of the UV background, baryonic feedback related to star-formation and SMBHs, and halos growing in mass and increasing their virial temperatures. Gas tends to remain at these hot temperatures $\sim10^{5.5-7}$~K from $z\sim2$ until $z\sim0.01$, before eventually dropping to the cool regime shortly before $z=0$.

Surprisingly, the roughly constant temperature reached shows no clear dependence with $z=0$ cluster mass. The progenitors of the most massive present-day clusters can show relatively low mean temperatures ($\sim10^{5.5}$~K) across all redshifts, while some low- and intermediate-mass cluster progenitors can have higher average temperatures ($\sim10^{7}$~K) in the tracers that form their eventual cool gas. This value does not represent the $z=0$ virial temperature, nor the increasing virial temperature of the main progenitor, but rather the spectrum of progenitor halos that will assembly at late times into the $z=0$ cluster. 

The median temperature evolves differently, peaking at temperatures of $\sim 10^5 \,\rm K$ at $z \sim 1$, before dropping to the cool regime, such that at $z = 0.1$ the median temperature is already $\sim 10^4\,\rm K$ (not shown). This discrepancy indicates that while at least half of this gas can already be cool at $z \sim 0.1$, the rest remains at high temperatures and thus dominates the mean until $z = 0.01$, when it rapidly cools in a few hundred million years to contribute to the present day cool gas component. 

Figure~\ref{fig:results:temperature_evo} also shows the evolution of the full temperature distributions of gas cells associated with selected tracers (right panel). Color denotes the mean tracer mass in each bin, per cluster. The division between the temperature regimes is indicated by two horizontal dashed white lines. In addition, the evolving dashed white lines shows the mean virial temperature of the cluster sample, while the two dotted lines bracket the minimum and maximum. We again see that most gas begins at $\sim 10^4\,\rm K$ at $z\gtrsim 8$. At redshift $z \approx 5$, a fraction of the gas begins to heat up, following the rise in virial temperature of the cluster progenitors. However, the distribution is extraordinarily broad: between $z = 3$ and $z = 0.1$ temperatures range from $10^4 \,\rm K$ to the virial temperature of the cluster. Instead of sitting near the current virial temperature of the progenitor halo, gas is instead spread approximately uniformly in temperature.

At $z \lesssim 1$, an increasing amount of gas collects at cool temperatures between $10^4$ and $10^{4.5}\,\rm K$. This cool gas becomes more and more prominent with decreasing redshift, reflecting a gradual cooling of warm and hot gas towards redshift zero. Nonetheless, a significant gas mass remains at intermediate and high temperatures all the way until $z = 0$, especially in the high temperature band above $10^7$\,K. By construction of the analysis, all gas at $z = 0$ is cool, as reflected in the narrow vertical black band on the left-side of this panel. The abundance of cold, star-forming gas at $10^3\,\rm K$ steadily increases with decreasing redshift, indicating that the cool cluster gas gas today is increasingly found in the star-forming phase.

Most individual clusters have temperature distributions reminiscent of the median, not mean, across the sample (not shown). Cool gas initially heats to a wide range of temperatures, avoiding the virial temperature of the cluster progenitor, and slowly cooling out of this warm-hot phase at $z \lesssim 1$. That is, while a majority of the cluster gas is near the virial temperature at all times, a majority of the cool cluster gas today was never at the virial temperature of the cluster progenitor. 
This reflects the spectrum of progenitor halos of lower masses to which this gas belongs (see below for more details). However, some clusters show strong, nearly flat plateaus around the cluster virial temperature containing significant mass fractions at all redshifts after its formation at around $z \sim 2$. This gas can remain for a substantial amount of time near constant temperature until shortly before redshift $z = 0$, where it cools to become part of the selected sample of tracers. 

A visual inspection of the distributions for individual clusters reveals that the subset of clusters exhibiting such a plateau in progenitor temperatures is the same subset of clusters that show a secondary accretion track in their distance distributions. These are both signposts of $z=0$ cool-core clusters, i.e. massive halos that possess large reservoirs of central cool gas.

\subsection{Accretion origin of cool gas in clusters}
\label{ssc:results:structural_origin_of_cool_gas}

\begin{figure*}
    \centering
    \includegraphics[width=0.49\textwidth]{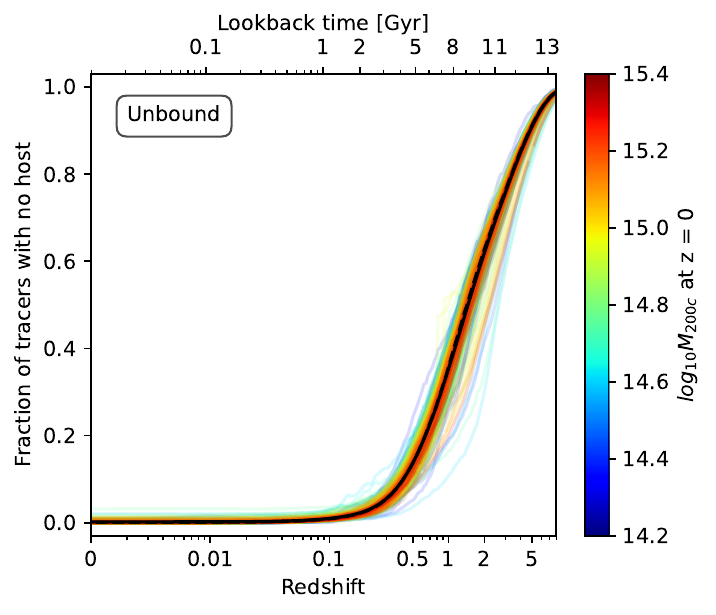}
    \includegraphics[width=0.49\textwidth]{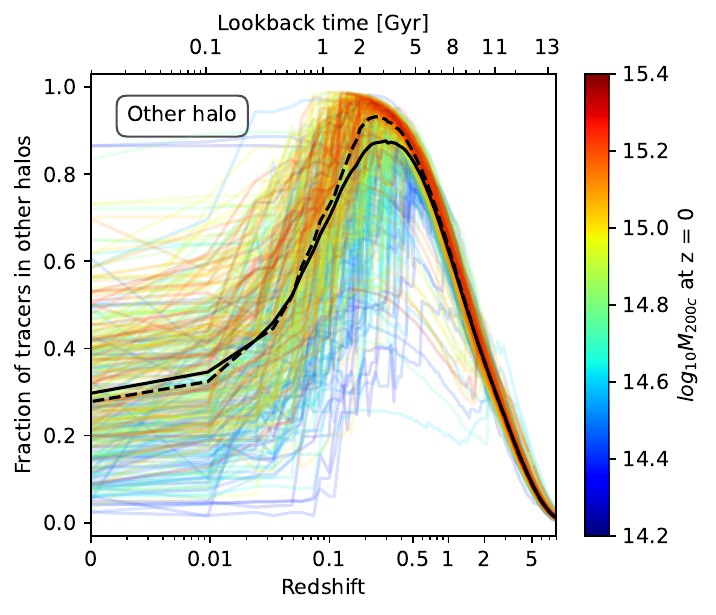}
    \includegraphics[width=0.49\textwidth]{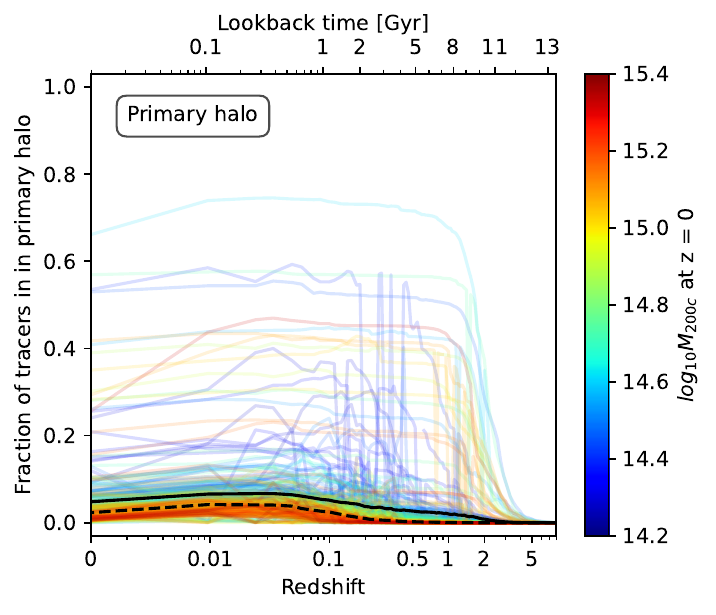}
    \includegraphics[width=0.49\textwidth]{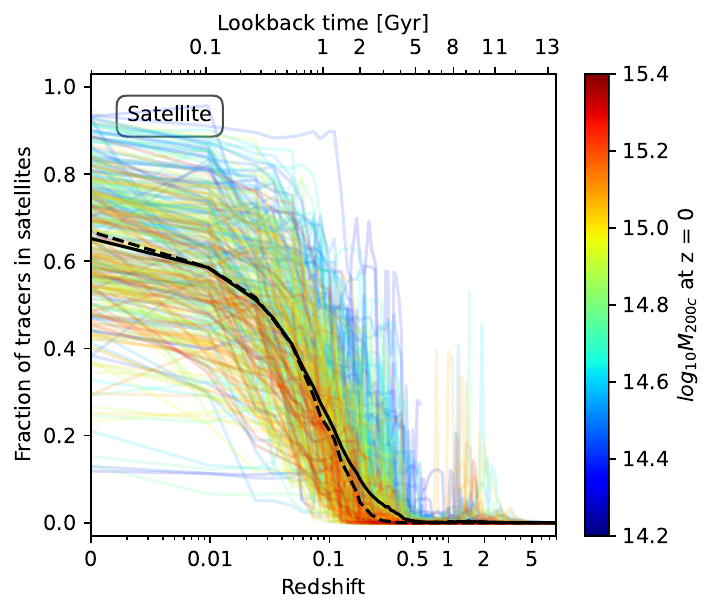}
    \caption{The origin of cool gas in clusters. Here we show the fraction of tracers that contribute to $z=0$ cool gas in clusters, from TNG-Cluster, in each of four origin categories, as a function of redshift. The categories are: `unbound' tracers that are outside of all halos (top left), `other halo' tracers that are within a halo other than the main progenitor of the $z=0$ cluster (top right), `primary halo' tracers that are within the central galaxy or diffuse ICM of the main progenitor (bottom left), and `satellite' tracers that are within satellite galaxies/subhalos of the main progenitor (bottom right). See Section~\ref{ssc:methods:analysis_choices}). Each halo of TNG-Cluster is shown with an individual line, color showing $M_{200c}$ at $z=0$, while the black solid (dashed) line shows the mean (median) trend. The majority of redshift $z=0$ cool gas is contributed to clusters by infalling satellite galaxies and merging of other halos.}
    \label{fig:results:parents}
\end{figure*}

What accretion channel is responsible for the $z=0$ cool gas in clusters? Figure~\ref{fig:results:parents} characterizes the fraction of tracers that belong to the four parent categories (as defined in Section~\ref{ssc:methods:analysis_choices}) as a function of redshift. These categories are: unbound (i.e., residing in the IGM, top left), in a halo other than the main progenitor (top right), in the central galaxy or diffuse halo of the main progenitor (bottom left), or in satellites of the main progenitor (bottom right). For brevity, we refer to these categories as unbound, other halo, primary, and satellite, respectively.
Colored lines represent individual clusters, with color indicating their $z=0$ halo mass. The black solid and dashed lines denote the mean and median values across all clusters, respectively.

The top left panel shows the fraction of tracers belonging to the diffuse IGM. It starts at near unity for all clusters at $z = 8$ and then monotonically drops to near zero. At $z = 0.1$, almost none of the cool gas predecessors are in the IGM anymore. There are only minor differences between clusters, with the mean and median lines almost identical, and individual halos likewise so. This process of cosmic cool gas accretion reaches half completion at $z \sim 2$ \citep[see][]{Wittig2025}.

The top right panel illustrates the fraction of tracers residing in other halos, regardless of their mass. Consequently, this category includes tracers in low-mass galaxies, group-sized halos, and even other protoclusters. Initially, at $z = 8$, this fraction is zero, but it increases to a maximum of $\approx 80-90\%$ by $z \sim 0.3$, after which it declines, reaching a mean fraction of approximately 30\% at redshift zero, although there a is a large halo-to-halo variation. The final fractions at $z=0$ also show a large spread, ranging from over 80\% to just a few percent. This spread reflects the clustering of nearby halos within $2R_{200c}$: while these halos are spatially close, they have not yet merged by $z=0$. No clear mass trend emerges, but at low redshifts, lower fractions of tracers in other halos are more common in low-mass clusters, whereas intermediate- to high-mass clusters are more likely to exhibit higher fractions.

The bottom left panel shows the fraction of tracers in the primary halo, either as part of its central galaxy or as part of the ICM. On average, this fraction remains low across all of cosmic time, below 10\% for the majority of clusters. In rare cases (5/352 clusters), however, it can reach up to 50\%.\footnote{Some trajectories show erratic behavior, bouncing between high and low values in adjacent snapshots. This is an artifact of the `switching' problem of the merger trees, but is subdominant, and does not correspond to any physical phenomenon.} The clusters that deviate strongly from the median curve are the same clusters with unusual distributions in distance (Figure~\ref{fig:results:tracers_dist}) and temperature (Figure~\ref{fig:results:temperature_evo}). Namely, these are clusters where a majority of the progenitor material of their $z=0$ cool gas components is in the primary halo of the cluster, and this cool gas tends to be located within the cluster cores.

\begin{figure*}
    \centering
    \includegraphics[width=0.49\textwidth]{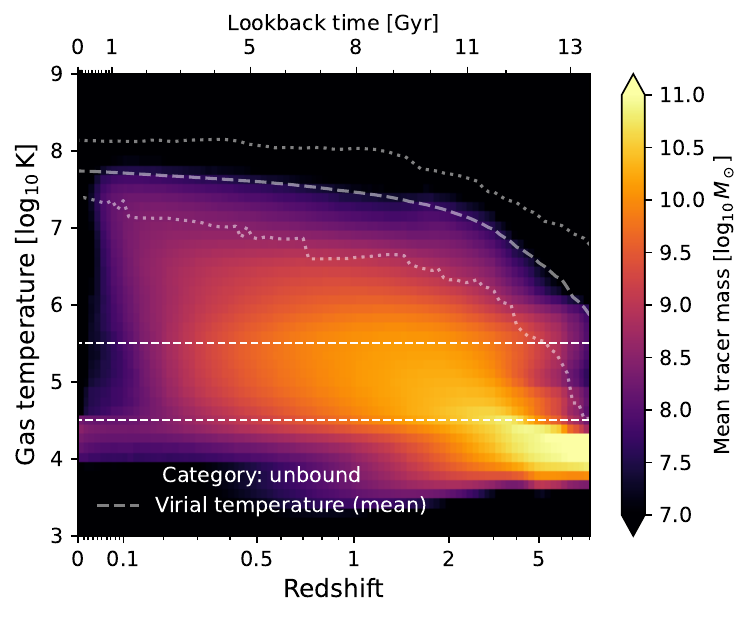}
    \includegraphics[width=0.49\textwidth]{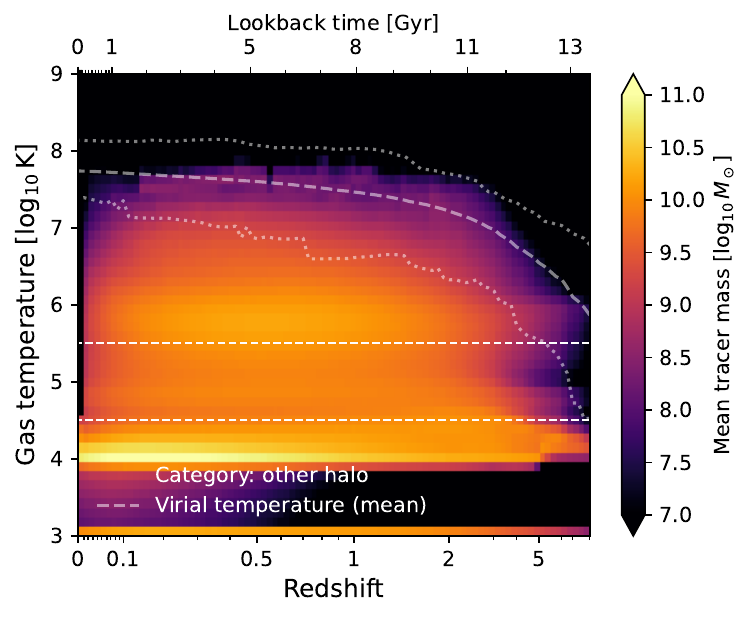}
    \includegraphics[width=0.49\textwidth]{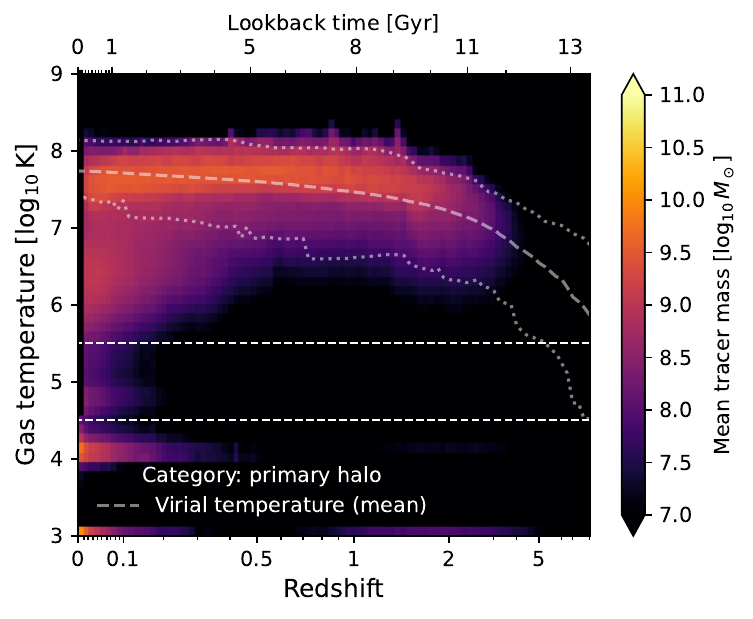}
    \includegraphics[width=0.49\textwidth]{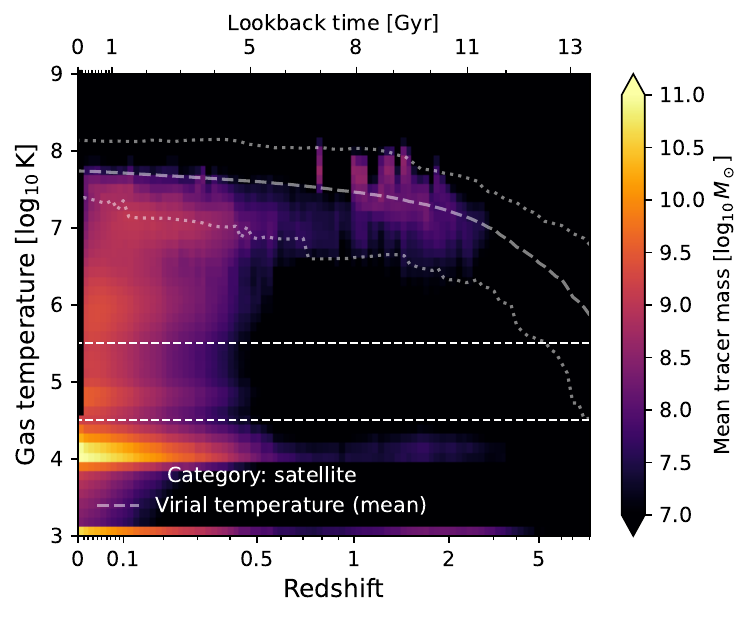}
    \caption{Time evolution of temperature distributions of selected tracers with redshift. As in Figure~\ref{fig:results:temperature_evo} (right panel), but splitting by parent categories (Section~\ref{ssc:methods:analysis_choices}). Each panel shows tracers of only one origin category. Top left: tracers outside of all halos i.e. in the IGM (unbound). Top right: Tracers in other halos i.e. not within the main progenitor of the primary cluster halo (other halo). Bottom left: Tracers in the primary cluster halo -- that is, the central galaxy or the diffuse ICM -- or its progenitors (primary halo). Bottom right: Tracers in satellites of the primary cluster halo (satellites). The particles in the primary halo contribute to redshift $z=0$ cool gas via in-situ cooling of virialized hot gas, while satellites and other halos contribute primarily pre-cooled gas to clusters.}
    \label{fig:results:temperature_2Dhists_split}
\end{figure*}

Lastly, the bottom right panel shows the fraction of tracers in satellite galaxies of the cluster. This fraction remains near zero for almost all clusters until $z \sim 0.5$, where it begins to rise. Lower mass clusters see an increase earlier than more massive clusters, reflecting that more massive clusters tend to assemble at later times. At $z < 0.1$, i.e. when the majority of cool gas tracers have entered $2R_{200c}$, this satellite fraction continues to rise. At redshift zero, the simulated clusters have diverse fractions of cool gas in satellites. On average, 65\% of cool gas is located in satellites at redshift at $z = 0$, but the spread is considerable, from 90\% to $<40$\% and even $<20$\%. In these latter cases, the fraction of tracers in the primary halo is large.

The other halo and satellite origins are linked: both reflect the accretion of pre-existing halos onto the cluster. Galaxies within these distinct, other halos eventually merge into the cluster and become satellites. In doing so, their tracers migrate from the top right to the bottom right panel, or to the bottom left panel if stripped out of the satellite. We consider whether the gas brought in by this accretion is already cool or must cool within the cluster below. Similarly, gas in the lower left panel i.e. the central galaxy or diffuse ICM can originate either from direct, cool accretion from the IGM \citep{Keres05}, cool gas stripped from satellites that is able to remain cool \citep[see][]{Rohr2023Jellyfish}, or the in-situ cooling of shock heated gas previously near the virial temperature \citep{Sharma2012Multiphase}. Overall, it is clear that the two right panels dominate, i.e. that for the vast majority of clusters, it is the accretion of substructure and eventually satellites that dominates the acquisition of their $z=0$ cool gas content.

Figure~\ref{fig:results:temperature_2Dhists_split} provides more detail through the time evolution of the full temperature distributions of gas associated with selected tracers. Each of the four panels shows tracers of one of the four parent categories, as above. As a result, we effectively decompose Figure~\ref{fig:results:temperature_evo} (right panel) into distinct origin channels.

\begin{figure*}
    \centering
    \includegraphics[width=0.5\textwidth]{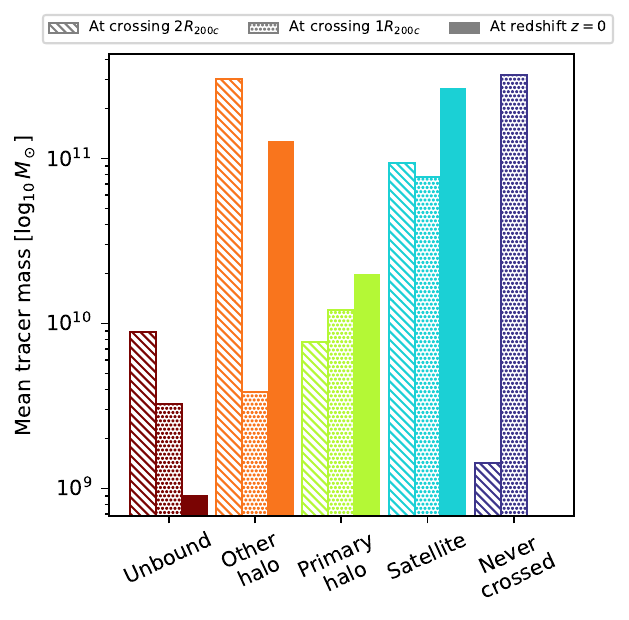}
    \includegraphics[width=0.95\textwidth]{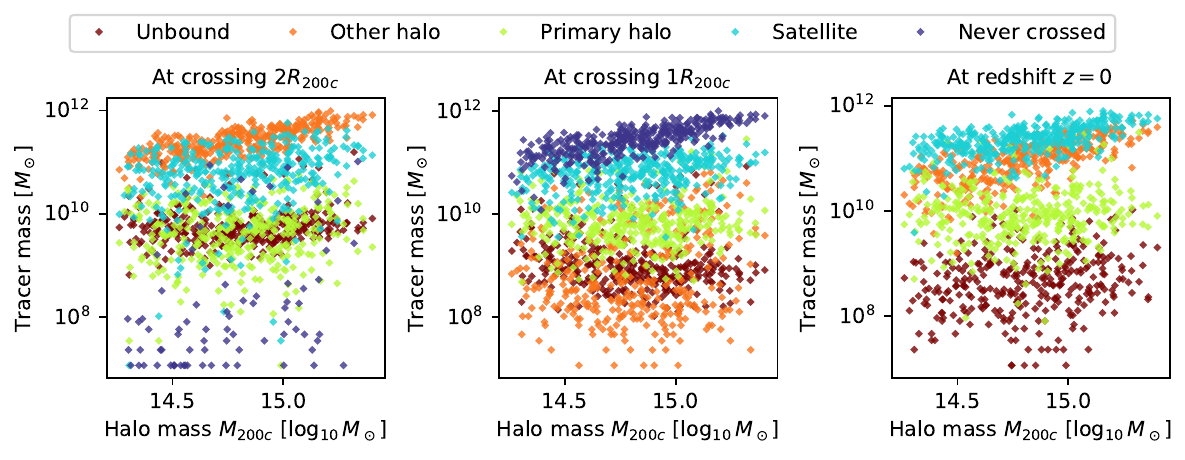}
    \caption{The fractional mass of cool gas in clusters by origin. Top panel: Mean mass of tracers per cluster in each parent category for all TNG-Cluster clusters at time of crossing $2R_{200c}$ (hatched), at crossing $R_{200c}$ (dotted), and at redshift $z = 0$ (solid). The parent categories are those defined in Section~\ref{ssc:methods:analysis_choices} and already characterized in Figures~\ref{fig:results:parents} and \ref{fig:results:temperature_2Dhists_split}. The rightmost blue bars additionally show the total mass of tracers that never crossed the respective distance, either because they remained outside this distance for all redshifts between $z = 8$ and today, or because they have constantly been within this radius since $z = 8$. For the redshift zero category, no such tracer exists, since all tracers are by definition in one of the four parent categories. Since all tracers have the same mass of $1.2 \times 10^7 \,\rm M_\odot$, the relative height of the bars gives the relative number of tracers as well. Bottom panel: Tracer mass in each origin category for individual clusters versus $M_{200c}$ of the respective cluster at $z=0$. The colors and respective categories are identical to the top panel. Each cluster is associated with five markers, one for each category. The majority of redshift $z=0$ cool gas predecessors is carried into the cluster $2R_{200c}$ region by satellites and other halos, in which it predominantly also remains until today, while $\sim 80\%$ of this gas never crosses the virial radius.}
    \label{fig:results:parent_category_bar_chart}
\end{figure*}

First, the top left panel shows the temperature distribution of only the unbound tracers i.e. of gas still in the diffuse IGM. This gas accounts for most of the structure at redshifts $z \gtrsim 2$, where the initial distribution of relatively cool gas is from the IGM. This material is then partially heated after accreting into halos of various masses, and potentially also by feedback effects. The amount of mass in this category decreases rapidly towards $z=0$ as the progenitor baryons of the clusters assemble first into smaller structures. While in this category, tracers do not heat up to the virial temperature of the cluster progenitors.

The top right panel shows tracers hosted by other halos, i.e. within gravitationally collapsed structure besides the proto-cluster halo itself. Towards lower redshifts, an increasing amount of this gas is in the cool and star-forming regime, but notable amounts of gas are found at higher temperatures up to the mean virial temperature of the clusters at almost all redshifts, with decreasing mean mass contributions towards higher temperatures. By $z = 0.1$, the majority of gas in this category is in the cool temperature regime. An excess of mass at $\sim 10^{5.5 - 6.0} \,\rm K$ extends from $z \sim 2$ to $z \sim 0.1$, the redshift at which most of the cool gas predecessors enter their host clusters $2R_{200c}$ radius. This region is relatively diffuse and arises due to virial shock heating in the lower mass halos that are being accreted by the clusters.

The bottom left panel shows history of the cool gas tracers within the primary halo. Little gas that is cool today was already within the cluster progenitor prior to $z\approx2$. Subsequently, the majority of this gas is hot, with temperatures near the virial temperature. This produces the plateau feature in the overall temperature distributions between $\sim 10^7 - 10^8\,\rm K$. A prominent tail forms extending to increasingly lower temperatures after $z \lesssim 0.6$, reflecting low-redshift, in-situ cooling of the hot ICM to temperatures below $10^{4.5}\,\rm K$ \citep{Sharma2012Multiphase,Voit2015Precipitation}. As we show here, it provides one origin channel for the cool gas content of $z=0$ clusters. However, this is only the case in a subset of halos. Inspection of the evolving temperature distributions for individual clusters shows that, in halos where such a feature is present, it contributes a large fraction of the $z=0$ cool gas. This drives the substantial increase in cool and star-forming gas since $z = 0.4$. However, cool gas brought in by satellites will also contribute to this category as soon as it is stripped during infall.

The bottom right panel shows the temperature distribution of gas that remains in (surviving) satellites. There is little such material at $z > 0.5$, indicating that satellites with cool gas preferentially accrete at late times, and/or that cool gas in cluster satellites is not a long-lived phenomenon \citep[see also][]{Rohr2023Jellyfish}. The patchy nature of the distribution at these redshifts reflects the discretized distribution of infalling satellites with cool gas within the cluster progenitor halo. Below redshift $z \sim 0.5$ however, this satellite gas spans a wide range of temperatures, from the star-forming temperature to the virial temperature. The majority contribution is in the cool regime, contributed both by cool gas around $10^4\,\rm K$ and by star-forming gas. This implies that most of selected tracers, entering the cluster $2R_{200c}$ region as part of satellite galaxies at $z \approx 0.1$, does so in an already cool state, likely within the ISM of the infalling satellites.

We conclude by developing an overall picture of these origin channels, and how they depend on halo mass. Figure~\ref{fig:results:parent_category_bar_chart} shows the relative importance of each of our four parent categories (top panel), from left to right: unbound, other halo, primary halo, and satellite. The subset of tracers that never cross $2R_{200c}$ or $R_{200c}$ are indicated by the final set of bars. In each of these five cases, we measure the total mass of tracers in that category, taking the average across the TNG-Cluster sample, at three times. Hatched bars show the $2R_{200c}$ crossing time, dotted bars show the $R_{200c}$ crossing time, and solid bars show the result at $z = 0$. 

When crossing $2R_{200c}$ (dashed bars), the smallest contributions arise from tracers in the IGM and within the primary progenitor halo already. The largest contribution by far comes from other halos, with satellites a close second. The distinction between these two is somewhat arbitrary, as an `other halo' becomes a satellite when it enters the halo, whose extent is not physical. Clearly, the accretion of the material into the $2R_{200c}$ regions of clusters that forms the $z=0$ cool gas in that same region is dominated by the accretion of substructure. A few particles have been within $2R_{200c}$ since $z=8$ and as such never crossed $2R_{200c}$ (blue bar).

When crossing $R_{200c}$ (dotted bars), the majority of tracers are part of satellites.\footnote{At the same time, the large blue dotted bar is larger, indicating that the majority of cool gas within our larger region of $2R_{200c}$ at $z=0$ have never been within the virial radius of the halo.} Other halos play only a minor role, and this is comparable to the contribution of tracers that are already within the ICM of the progenitor. This is somewhat surprising, as we might expect a larger amount of material to become part of the primary halo before crossing its virial radius. Finally, the total mass of tracers in the unbound state is lower when crossing the virial radius than when crossing $2R_{200c}$, with most gas having entered some structure by this distance. 

At $z=0$ all the cool gas predecessors have by definition cooled to below the threshold of $10^{4.5}\,\rm K$ and therefore the mass in each category represents the present-day cool gas mass. Unbound particles only play a role in the outskirts near $2R_{200c}$ and contribute the smallest fraction. Cool gas belonging to the primary halo contributes more than an order of magnitude more mass. The majority of cool gas at $z=0$ however is situated in other halos in the outskirts of the cluster, and in satellite galaxies. This signifies that most cool gas at redshift zero has not yet been removed from the satellites it came in with, and has not yet been incorporated into the primary halo. 

The bottom three panels of Figure~\ref{fig:results:parent_category_bar_chart} show the contributions of each parent category to each individual cluster, as a function of $z=0$ halo mass. The left panel shows the values when crossing the $2R_{200c}$ sphere, the middle panel when crossing the virial radius, and the right panel at $z = 0$. In general, the vast majority of the cluster population is consistent with the (mean) trends discussed above, with some notable outliers.

Upon entering $2R_{200c}$ (left panel), the contribution from satellite galaxies is highly variable, spanning several orders of magnitude from halo to halo. This variability reflects the scatter in environment and richness within the TNG-Cluster sample. The contributions from the other halo and satellite categories increase with halo mass, mirroring the general trend of cool gas mass scaling with cluster mass (see Figure~\ref{fig:results:cool_gas_vs_cluster_mass}). In contrast, the other categories show much weaker dependence on halo mass.
When crossing the virial radius (middle panel), the contribution from still distinct other halos is generally low but exhibits significant scatter. Additionally, positive mass trends for the dominant categories become evident in this panel.

Finally, the structure of cool gas at $z=0$ (right panel) shows satellite galaxies and other halos as the main origin for cool gas. However, there are clusters which show comparable gas masses in their primary halo: these are the halos with strong in-situ cooling of their hot ICM. While the contributions by satellites and other halos show a clear trend with cluster mass, no such trend is apparent from the gas in the primary halo, which leads to lower mass clusters containing on average a higher fraction of cool gas in their primary halo. This is consistent with the decreasing abundance of cool-core clusters with increasing halo mass in TNG-Cluster at $z=0$ \citep{Lehle2024CoolCores}.

The overall diversity visible across each of these five categories, and its evolution as material accretes from the cluster outskirts into its center, shows that a variety of assembly histories and physical processes contribute to the $z=0$ cool gas content of galaxy clusters.


\section{Conclusions} \label{sec:conclusions}

In this work, we have quantified the abundance, spatial distribution, and origin of cool ($\leq 10^{4.5} \,\rm K$) gas in 632 simulated clusters at $z=$ with masses $M_{200c} \sim 10^{14-15.4}\,\rm{M_\odot}$. To do so, we have combined the TNG-Cluster and TNG300 cosmological magnetohydrodynamical simulations and studied the clusters gas out to twice their virial radius, accounting for both diffuse ICM as well as gas in and around cluster galaxies. By using Monte-Carlo Lagrangian tracer particles, we could identify the origin channels of this cool gas and quantify the contributions from diffuse accretion, mergers, and in-situ cooling processes. Our key findings are as follows:

\begin{itemize}
    \item At $z = 0$, cool gas is present in nearly all clusters in the sample, with mass fractions within $2R_{200c}$ ranging from a few percent down to $\sim 10^{-4}$. The cool gas fraction decreases with increasing cluster mass. However, the total cool gas mass within $2R_{200c}$ remains substantial, ranging from $\sim 10^{10} \,\rm M_\odot$ to $\sim 10^{12} \,\rm M_\odot$, and increases with cluster mass (Figure~\ref{fig:results:mass_trends_tng300}).
    \item The cool gas content in clusters exhibits weak correlations with many halo and galaxy properties (Table~\ref{tb:results:cluster_properties}). However, strong positive correlations are found with the star formation of the central and of all cluster galaxies (Figures~\ref{fig:results:cool_gas_vs_cluster_mass} and ~\ref{fig:results:cool_gas_vs_cluster_mass_core_only}).
    \item Outside of cluster cores, halo outskirts typically host the majority of the cool gas. On average, only $\sim 20\%$ of the cool gas mass is located within $R_{200c}$ at $z = 0$, though in some cases, this fraction can be as high as 80\% (Figure~\ref{fig:results:radial_dependence_temperature}). The spatial distribution of this gas is clumpy, as it is predominantly found within cluster galaxies (Figure~\ref{fig:results:single_halo_vis}).
    \item In cluster cores ($< 0.05 R_{200c}$), while most halos contain no cool gas, a subset of the cluster population does. The central cool gas mass ranges from $\sim 10^7\,\rm M_\odot$ to $\gtrsim 10^{11}\,\rm M_\odot$ (Figure~\ref{fig:results:radial_dependence_temperature_core}). Within 1\% of the virial radius, the mean cool gas density is comparable to the total gas density (Figure~\ref{fig:results:radial_density_profiles}).
    \item Cool gas is predominantly inflowing throughout the cluster environment, with velocities on the order of the cluster virial velocity (Figures~\ref{fig:results:radial_density_profiles} and \ref{fig:results:velocity_distribution}).
    \item The baryons that give rise to the $z=0$ cool gas in clusters typically enter halos around $z \sim 0.1$. However, in some halos, this material is accreted as early as $z \sim 2$ (Figure~\ref{fig:results:tracers_dist}). 
    \item Similarly, this gas first cools at $z \sim 2$, but only cools for the final time by $z \sim 0.1$. This suggests that most of the gas experiences multiple heating and cooling cycles before settling into its final cooled state (Figure~\ref{fig:results:cooling_times}). The evolving temperature distribution of this material indicates that virial shock heating usually occurs in halos less massive than the main progenitor of the protocluster (Figure~\ref{fig:results:temperature_evo}).
    \item The majority of the gas that eventually forms the $z=0$ cool gas in clusters predominantly enters these halos in a pre-cooled state, primarily through the accretion of satellites and other halos. On average, these sources contribute approximately $\sim 65\%$ and $\sim 30\%$, respectively, though the exact fractions vary across clusters (Figures~\ref{fig:results:parents} and \ref{fig:results:parent_category_bar_chart}).
    \item A subset of the $z=0$ cluster population exhibits significant in-situ cooling of the hot ICM. This cooling occurs on short timescales and contributes substantially to the late-time cool gas component of these clusters, where it is typically found in the centers of cool-core clusters (Figure~\ref{fig:results:temperature_2Dhists_split}).
\end{itemize}

Our analysis has quantified the abundance and properties of cool gas in simulated galaxy clusters at $z=0$, while also providing a theoretical framework for its origin. Future cross-simulation comparisons will help determine whether our findings are specific to TNG-Cluster and the TNG model or if they hold more generally across cosmological simulations. Additionally, direct comparisons with observations of cool gas in $z \sim 0$ clusters and high-redshift protoclusters will not only provide unique constraints on the simulations but also offer new interpretative insights into the observational data itself.

\section*{Data Availability}

The IllustrisTNG simulations themselves are publicly available and accessible at \url{www.tng-project.org/data} \citep{Nelson2019a}, where also the TNG-Cluster simulation is now made public. Data directly related to this publication is available on request from the corresponding author. This work has benefitted from the \texttt{scida} analysis library \citep{Byrohl2024Scida}.

\begin{acknowledgements}

DN and MA acknowledge funding from the Deutsche Forschungsgemeinschaft (DFG) through an Emmy Noether Research Group (grant number NE 2441/1-1). MA is supported at the Argelander Institute f\"ur Astronomie through the Argelander Fellowship. This work is supported by the Deutsche Forschungsgemeinschaft (DFG, German Research Foundation) under Germany's Excellence Strategy EXC 2181/1 - 390900948 (the Heidelberg STRUCTURES Excellence Cluster). AP acknowledges funding from the European Union (ERC, COSMIC-KEY, 101087822, PI: Pillepich).

The TNG-Cluster simulation suite has been carried out with compute time awarded under the TNG-Cluster project on the HoreKa supercomputer, funded by the Ministry of Science, Research and the Arts Baden-Württemberg and by the Federal Ministry of Education and Research; the bwForCluster Helix supercomputer, supported by the state of Baden-Württemberg through bwHPC and the German Research Foundation (DFG) through grant INST 35/1597-1 FUGG; the Vera, MCobra, and Raven clusters of the Max Planck Computational Data Facility (MPCDF); and the BinAC cluster, supported by the High Performance and Cloud Computing Group at the Zentrum für Datenverarbeitung of the University of Tübingen, the state of Baden-Württemberg through bwHPC and the German Research Foundation (DFG) through grant no INST 37/935-1 FUGG. The three original TNG simulations were run with compute time awarded by the Gauss Centre for Supercomputing (GCS) under GCS Large-Scale Projects GCS-ILLU and GCS-DWAR on the Hazel Hen supercomputer at the High Performance Computing Center Stuttgart (HLRS). The calculations in this paper were carried out on the Vera cluster of the Max Planck Institute for Astronomy (MPIA).
\end{acknowledgements}

\bibliographystyle{aa}
\bibliography{references}

\end{document}